\documentclass{article}

\usepackage{arxiv}
\usepackage[utf8]{inputenc} 
\usepackage[T1]{fontenc}    
\usepackage{hyperref}       
\usepackage{url}            
\usepackage{booktabs}       
\usepackage{amsfonts}       
\usepackage{nicefrac}       
\usepackage{microtype}      
\usepackage{lipsum}
\usepackage{natbib}

\usepackage{eqnarray,amsmath}
\usepackage{algorithm}
\usepackage{graphicx}
\usepackage{multirow}

\usepackage[shortlabels]{enumitem}

\newcommand{\vect}[1]{\boldsymbol{#1}}

\newcommand{\ess}{\mbox{ESS}}
\newcommand{\E}{\mathsf{E}}
\renewcommand{\Pr}{\mathsf{P}}
\newcommand{\dNorm}{\mathcal{N}}
\newcommand{\dGam}{\mathcal{G}}

\newcommand{\med}{\mathrm{median}}

\DeclareMathOperator*{\argmin}{arg\,min}
\DeclareMathOperator*{\argmax}{arg\,max}

\newtheorem{prop}{Proposition}

\title{Accelerating sequential Monte Carlo with surrogate likelihoods}

\author{
 Joshua J. Bon\\
  School of Mathematical Sciences\\
  Queensland University of Technology\\
  \texttt{joshuajbon@gmail.com} \\
\And
 Anthony Lee \\
  School of Mathematics\\
  University of Bristol\\
  \texttt{anthony.lee@bristol.ac.uk} \\
\And
  Christopher Drovandi\\
  School of Mathematical Sciences\\
  Queensland University of Technology\\
  \texttt{c.drovandi@qut.edu.au} \\
}

\begin{document}
\maketitle

\begin{abstract}
Delayed-acceptance is a technique for reducing computational effort for Bayesian models with expensive likelihoods. Using a delayed-acceptance kernel for Markov chain Monte Carlo can reduce the number of expensive likelihoods evaluations required to approximate a posterior expectation. Delayed-acceptance uses a surrogate, or approximate, likelihood to avoid evaluation of the expensive likelihood when possible. Within the sequential Monte Carlo framework, we utilise the history of the sampler to adaptively tune the surrogate likelihood to yield better approximations of the expensive likelihood, and use a surrogate first annealing schedule to further increase computational efficiency. Moreover, we propose a framework for optimising computation time whilst avoiding particle degeneracy, which encapsulates existing strategies in the literature. Overall, we develop a novel algorithm for computationally efficient SMC with expensive likelihood functions. The method is applied to static Bayesian models, which we demonstrate on toy and real examples, code for which is available at \url{https://github.com/bonStats/smcdar}.
\end{abstract}


\acknow{JJB is a recipient of a PhD Research Training Program scholarship from the Australian Government. JJB, AL, and CD thank the Australian Research Council (ARC) Centre of Excellence for Mathematical and Statistical Frontiers for financial support (CE140100049). AL and CD were supported by an ARC Research Council Discovery Project (DP200102101). AL was supported by an EPSRC grant (EP/R034710/1) and received travel funding from the Statistical Society of Australia. JJB and CD also thank the Centre for Data Science at QUT for support.}

\newpage

\section{Introduction}
\label{sec:intro}
A significant barrier for statisticians and practitioners is the computational requirements for handling complex statistical models with challenging likelihood functions. Particularly in Bayesian modelling, computationally expensive likelihoods can impede inference and cause drastic reductions in accuracy on a limited computational budget.

Bayesian statistics is becoming more popular in a growing number of disciplines due to its principled framework for uncertainty quantification in parameter estimation, model selection, and prediction. To keep up with practitioners desire to develop and use more sophisticated and realistic stochastic models, there is a demand for improved statistical methodology for parameter estimation and model selection from data.

This article focuses on sequential Monte Carlo \citep[SMC,][]{chopin2002sequential,del2006sequential}, a theoretically justifiable method for Bayesian parameter estimation and model selection. SMC works by propagating a set of weighted samples (called particles) through a sequence of distributions connecting a simple distribution to a more complex distribution that is difficult to sample from directly. SMC for Bayesian inference typically has the posterior distribution as its final target.

The basic ingredients of SMC are reweighting, resampling, and mutation of the particles. SMC can be appealing for a wide variety of inference problems as it can handle multimodality, is easily parallelisable as well as adaptable, and provides an estimate of the posterior normalising constant (the evidence, which can be useful for model selection). Despite the widespread success of SMC, it can involve a large number of likelihood evaluations, which is expensive for models with computationally intensive likelihood functions.

An alternative and popular method for Bayesian sampling is Markov chain Monte Carlo (MCMC), which involves constructing a Markov chain with the target posterior as its limiting distribution. The cornerstone of MCMC, the Metropolis-Hastings (MH) algorithm \citep{metropolis1953equation,hastings1970monte}, also suffers computationally when the likelihood function is expensive. An alternative MCMC kernel, delayed-acceptance (DA), can be used to alleviate some of this computational burden \citep{fox1997sampling, christen2005markov}. This method uses a cheap surrogate of the target to perform an initial screening phase of each proposed parameter value in MCMC. It resembles the MH kernel, but with two stages.

 In this paper, we use the terms \textit{surrogate likelihood} and \textit{surrogate posterior} to refer to a particular approximate likelihood and posterior, respectively. Using the DA kernel, if the proposal is a poor candidate for the target distribution, according to the surrogate, it will have a high probability of being rejected without needing to evaluate the expensive likelihood. If the surrogate can accurately predict which proposals are likely to be accepted or rejected using the actual likelihood, a more efficient MCMC algorithm can be expected. Issues can arise when the tails of the surrogate are mismatched to the target, but there are adjustments that can be made to address this \citep{banterle2019accelerating}.

For DA-MH to be successful, the surrogate must be (i) relatively cheap and (ii) roughly proportional to the posterior as a function of the parameter. In some applications, a cheap version of the model may be directly available, such as the linear noise approximation for Markov processes \citep{elf2003fast,stathopoulos2013markov}. In other applications, a general surrogate can be developed using ideas from emulation, for example regression trees \citep{sherlock2017adaptive} and Gaussian processes \citep{conrad2016accelerating,drovandi2018accelerating}. In the former scenario, the cheap approximation may lack sufficient flexibility to satisfy (ii). In the latter scenario, the training region for the emulator can be hard to identify and again might not be flexible enough to satisfy (ii).

The aim of this paper is to develop novel delayed-acceptance methods that are efficient and automated by harnessing the SMC framework. DA has been leveraged in SMC previously in the setting of approximate Bayesian computing \citep{everitt2017delayed}. We propose surrogate accelerated SMC to improve delayed-acceptance, and use of surrogate likelihoods in SMC, through three avenues, which also constitute the main contributions of this work. Firstly, delayed-acceptance can be used directly in the mutation step of the SMC algorithm, reducing costly evaluations of the likelihood. Secondly, the particles can be used to tune the surrogate likelihood to better match the full likelihood. Lastly, we can change the annealing strategy of SMC to only use the surrogate likelihood at first, then anneal to the full posterior which then uses the full likelihood. These adaptations are made possible from access to the population of particles in SMC.

In creating our new SMC sampler, we also develop a framework for adaptively minimising the computation time accrued in the mutation step of SMC whilst achieving a minimum level of particle diversification. This framework encompasses existing strategies in the literature with appropriate approximations.

The paper proceeds as follows. Section~\ref{sec:smc-intro} provides an overview of SMC, whilst Section~\ref{sec:da-mh-intro} introduces the delayed-acceptance Metropolis-Hastings algorithm. In Section~\ref{sec:tuning-mh-smc} we discuss existing tuning strategies for mutation kernels in SMC and propose a framework for such strategies in the context of MH kernels. Section~\ref{sec:tuning-da-mh} extends our tuning framework to delayed-acceptance in SMC, and discusses the other ways surrogate likelihoods can be utilised for improved computationally efficiency. Section~\ref{sec:sim-ex} contains a simulation study investigating the effect of computation cost differentials between the surrogate and full likelihood functions, as well as different tuning parameters and strategies. Section~\ref{sec:whittle} applies our methods to times series models using the Whittle likelihood approximation \citep{whittle1953estimation}.

\section{Sequential Monte Carlo}
\label{sec:smc-intro}
Sequential Monte Carlo generates samples from a sequence of distributions forming a bridge between a simple initial distribution, that is easy to sample from, and a final target distribution. The target distribution is generally the posterior distribution in Bayesian analysis.  In this paper we consider a power-likelihood path connecting the prior, or other starting distribution, to the posterior distribution. As such, we use this temperature annealed sequence to illustrate SMC methodology, which takes the form
	\begin{align*}
	p_t(\vect{\theta}|\vect{y}) &\propto p(\vect{\theta}|\vect{y})^{\gamma_t} p_0(\vect{\theta})^{1-\gamma_t}, \\
	p(\vect{\theta}|\vect{y}) &\propto p(\vect{y}|\vect{\theta})\pi(\vect{\theta})
	\end{align*}
where $p(\vect{\theta}|\vect{y})$ is the posterior density, $p(\vect{y}|\vect{\theta})$ is the likelihood implied by a statistical method assumed to derive observed data $\vect{y}$ and is parameterised by $\vect{\theta}$, $\pi(\vect{\theta})$ is the prior distribution, and $0 = \gamma_0 < \gamma_1 < \cdots < \gamma_T = 1$. The ultimate target distribution is the posterior $p_T(\vect{\theta}|\vect{y}) \equiv p(\vect{\theta}|\vect{y})$.  The sequence of targets begins with the initial distribution $p_0(\vect{\theta})$ when the annealing parameter is $\gamma_t = 0$ and transitions to the target posterior by steadily increasing $\gamma_t$ and ultimately terminates at $\gamma_T = 1$. It is often the case that the prior is chosen to be the initial distribution by taking $p_0(\vect{\theta}) = \pi(\vect{\theta})$, resulting in likelihood tempering of the form $p_t(\vect{\theta}|\vect{y}) \propto p(\vect{y}|\vect{\theta})^{\gamma_t} \pi(\vect{\theta})$.

At each iteration of the SMC algorithm, the $t$th target distribution $p_t$ is represented by a weighted empirical measure
	\begin{align*}
	\hat{p}^{N}_{t} &= \sum_{i=1}^N W_t^i \delta_{\vect{\theta}_t^i}(\vect{\theta})
	\end{align*}
using $N$ particles, $\vect{\theta}_{t}^{i}$, each associated with weights $W_t^i$, such that ${\scriptstyle\sum_{i=1}^{N}} W_t^i = 1$. The empirical measure can be used to approximate expectations with respect to $p_t$, and ultimately the posterior.

As stated, an SMC algorithm iterates through reweighting, resampling and mutation steps to migrate the population of particles from $p_0(\vect{\theta})$ to $p(\vect{\theta}|\vect{y})$.  For static Bayesian models under posterior tempering, the reweighting step uses the current particles and weights from $\hat{p}^{N}_{t-1}$, to generate the next set of particles and weights for $\hat{p}^{N}_{t}$. The new weights are
	\begin{equation*}
		W_{t}^i \propto w_{t}^i = W_{t-1}^i  \frac{p_{t}(\vect{\theta}_{t-1}^i| \vect{y})}{p_{t-1}(\vect{\theta}_{t-1}^i| \vect{y})} \label{eq:reweight-weight}
	\end{equation*}
	and the updated particles and normalised weights are given by
	\begin{equation*}
	\vect{\theta}_{t}^i = \vect{\theta}_{t-1}^i, \quad W_{t}^i = \frac{w_{t}^i}{\sum_{k=1}^{N}w_{t}^k}
	\end{equation*}
for $i=1,\ldots,N$.

We can measure the quality of the SMC sample at iteration $t$ via the effective sample size (ESS).  The ESS is often estimated via the normalised weights
    \begin{equation}
        \ess_t = 1/\sum_{i=1}^N (W_t^i)^2. \label{eq:ess}
    \end{equation}
When the ESS drops below some threshold $S$ (often set at $N/2$) an intervention is required to prevent degeneracy in the particle set.  Resampling $N$ particles from the particle set $\{\vect{\theta}_{t}^i\}_{i=1}^N$ with weights $\{W_{t}^i\}_{i=1}^{N}$ produces an unweighted empirical measure ($W_{t}^i = 1/N$) which also approximates the current tempered distribution. The resampling process duplicates particles with relatively high weights and drops those with relatively small weights so that the SMC algorithm can better explore high probability regions of the posterior. A number of resampling strategies are available including multinomial, stratified, or residual resampling \citep{kitagawa1996monte,liu1998sequential}.

In practice, $\ess_{t}(\gamma_{t})$ only depends on the next annealing parameter $\gamma_{t}$ since the particle values, $\vect{\theta}_{t}$, are fixed during the reweighting stage. As such, we can select $\gamma_{t}$ adaptively via the bisection method bounded by $(\gamma_{t-1},1]$ such that $\ess_{t}(\gamma_{t}) \approx S$, where $S$ is the targeted ESS \citep{jasra2011inference,beskos2016convergence}. If such a strategy is chosen then the particles are resampled every iteration.

To diversify the resampled particle set (which is now likely to contain duplicates), the particles can be perturbed with an MCMC kernel with invariant distribution $p_{t}$.  For simplicity, and consistency with other approaches in the literature \citep[see][for example]{chopin2002sequential,jasra2011inference,beskos2016convergence}, a Metropolis-Hastings kernel is used with proposal distribution, $q_{h}$, given by a multivariate normal (MVN) random walk
	\begin{align}
	q_{h}(\vect{\theta}_{*}|\vect{\theta}_{t}) &= \dNorm(\vect{\theta}_{*}; \vect{\theta}_{t}, h^2 \Sigma), \label{eq:mvn-mh-kernel}
	\end{align}
where, in this case, the proposal distribution has step size, $h$, as a tuning parameter. For other kernels however, a vector of tuning parameters may be appropriate. The covariance matrix $\Sigma$ can be set to the sample covariance estimated from the weighted particles before resampling, a choice adopted in our demonstrations in Sections~\ref{sec:sim-ex} and~\ref{sec:whittle}.

A proposed update, or proposal $\vect{\theta}_{*}^{i}$, is made for the $i$th particle, $\vect{\theta}_{t}^{i}$, by drawing a random variable according to \eqref{eq:mvn-mh-kernel}. In general, the proposal is accepted as the new value of the particle with probability
\begin{align*}
    \alpha(\vect{\theta}_{*},\vect{\theta}_{t}) = \min \left\{1, r(\vect{\theta}_{*},\vect{\theta}_{t}) \right\}, \text{ with } r(\vect{\theta}_{*},\vect{\theta}_{t}) = \frac{p_{t}(\vect{\theta}_{*}|\vect{y}) q_{h}(\vect{\theta}_{t} |\vect{\theta}^*) }{p_{t}(\vect{\theta}_{t}|\vect{y}) q_{h}(\vect{\theta}^*|\vect{\theta}_{t})},
\end{align*}
where, in this case, $r$ can be simplified to $\nicefrac{p_{t}(\vect{\theta}_{*}|\vect{y}) }{p_{t}(\vect{\theta}_{t}|\vect{y})}$ due to the symmetry of the MVN proposal distribution. If the proposal is rejected, the value of the particle remains unchanged. For a generic mutation kernel, $K_{\vect{\phi}}$($\vect{\theta}, \cdot)$, we will refer to the tuning parameters as the vector $\vect{\phi}$ henceforth.

It is often the case that one iteration of an MCMC kernel is not sufficient to diversify the particle set adequately. One can choose the kernel to be a cycle of primitive MCMC kernels, and it is often the case that SMC algorithms apply a given kernel multiple times for the mutation step. Whilst the number of repeats can be fixed, there are several adaptive methods for tuning the number of cycles required. Such adaptive methods can be informed by a pilot run of the mutation step \citep{drovandi2011likelihood,salomone2018unbiased}, or evolve based on probabilistic rules \citep{fearnhead2013adaptive}. We better explore these tuning methods in Sections~\ref{sec:tuning-mh-smc} and~\ref{sec:tuning-da-mh} where we present our framework for tuning kernels based on minimal diversification and computation time.

\begin{algorithm}
    \caption{Adaptive resample-move sequential Monte Carlo for static Bayesian models\label{alg:SMC-alg}}
    \textbf{Input:} Number of particles, $N$; Initial distribution, $p_{0}$; Family of MCMC kernels, $K_{\vect{\phi}}$; Kernel tuning set,~$\Phi$; ESS threshold, $S$; Maximum mutation steps, $M$.

    \begin{enumerate}
        \item Initialisation.
        \begin{enumerate}[(a)]
            \item $t=0$, $\gamma_0 = 0$, and $W_{0}^{i} = \frac{1}{N}$ for $i\in \{1,\ldots,N\}$
            \item Simulate $\vect{\theta}_{0}^{i} \sim p_0$ for $i\in \{1,\ldots,N\}$ (\textit{perhaps from a previous SMC run})
        \end{enumerate}
        \item While $\gamma_t < 1$ increment $t$, then iterate through
        \begin{enumerate}[(a)]
            \item Calculate $\gamma_t = \sup_{\gamma \in (\gamma_{t-1}, 1]}\{\text{ESS}_{t}({\gamma}) \geq S\}$, the new temperature using the bisection method
            \item Compute new weights $w_t^i = W_{t-1}^i \frac{p_{t}(\vect{\theta}_{t-1}^i| \vect{y})}{p_{t-1}(\vect{\theta}_{t-1}^i| \vect{y})}$ then normalise $\check{W}_t^i = \frac{w_t^i}{\sum_{k=1}^{N}w_t^k}$ for $i\in \{1,\ldots,N\}$
            \item Resample particle $\vect{\theta}_{t}^i \sim \sum_{i=1}^{N}\check{W}_t^i\delta_{\vect{\theta}_{t-1}^i}(\cdot)$ then set $W_{t}^{i} = \frac{1}{N}$ for $i\in \{1,\ldots,N\}$
			\item  Tune the mutation kernel, $K_{\vect{\phi}}$, by selecting $\vect{\phi}^{\ast} \in \Phi$ in two steps, (d1) and (d2) (\textit{details in subsequent sections})
            \item For $s = 1,2,\ldots,M$ mutate particle $\check{\vect{\theta}}_{s}^i \sim K_{\vect{\phi}^{\ast}}$($\check{\vect{\theta}}_{s-1}^i, \cdot)$ for $i\in \{1,\ldots,N\}$

            \item Update particle $\vect{\theta}_{t}^i = \check{\vect{\theta}}_{M}^i$ for $i\in \{1,\ldots,N\}$
        \end{enumerate}
    \end{enumerate}
    \textbf{Output:} Particles $\{\vect{\theta}_{T}^i\}_{i=1}^{N}$ such that $\gamma_{T} = 1$. \vspace{0.2cm}\\
\end{algorithm}

We present a generic description of an adaptive SMC for static Bayesian models in Algorithm \ref{alg:SMC-alg}. This algorithm summarises the methodology outlined in this section for a particular instance of SMC, a resample-move algorithm \citep{gilks2001following} with power-likelihood annealing. Using the resample-move algorithm as a prototype simplifies the exposition of this paper, however the developments we make are applicable to a much wider class of SMC algorithms. The mutation kernel, $K_{\vect{\phi}}$($\vect{\theta}, \cdot)$, in Algorithm \ref{alg:SMC-alg} will typically be the Metropolis-Hastings kernel, or the delayed-acceptance kernel discussed in Section~\ref{sec:da-mh-intro} in the case of expensive likelihoods.


\section{Delayed-acceptance Metropolis-Hastings}
\label{sec:da-mh-intro}

Expensive likelihoods can considerably slow down Bayesian analysis when using computational methods for posterior approximation. In Metropolis-Hastings routines, one remedy is delayed-acceptance which uses a surrogate likelihood to first test whether a proposal is worthy of evaluation on the expensive full likelihood. If the surrogate is a good approximation to the full likelihood then this can be used to reject poor proposals early and avoid using unnecessary computation time evaluating the full likelihood.

Delayed-acceptance has its origins in Bayesian conductivity imaging \citep{fox1997sampling} followed by a more complete account in \citet{christen2005markov} who formalised the ergodic correctness of the DA-MH algorithm. More recently, \citet{banterle2019accelerating} proposed a modified version of delayed-acceptance which is robust to poor tail-coverage by the surrogate likelihood. They explore bounds for the variance of delayed-acceptance compared to standard Metropolis-Hastings. We use this version of delayed-acceptance in our applications of SMC.

Delayed-acceptance has been used to speed up a number of costly MCMC algorithms including pseudo-marginal methods \citep{golightly2015delayed, wiqvist2018accelerating}, approximate Bayesian computing \citep[ABC,][]{everitt2017delayed}, and Bayesian inverse problems \citep{cui2011bayesian}. It has also been combined with data subsampling \citep{quiroz2018speeding} and consensus MCMC \citep{payne2018two} to produce more general algorithms for accelerated MCMC. It is worth noting however, whilst delayed-acceptance targets the correct stationary distribution, not all of the aforementioned methods share this property after further approximations are made.

The surrogate likelihood in delayed-acceptance can be a deterministic approximation, such as the Linear Noise Approximation (LNA) for stochastic kinetic models \citep{golightly2015delayed}, but need not be. In particular \citet{sherlock2017adaptive} and \citet{wiqvist2018accelerating} use adaptive non-parametric approximations to the full likelihood. The former uses a nearest-neighbour approximation whilst the latter use a Gaussian process for the log-likelihood, in a similar manner to \cite{drovandi2018accelerating}.

More theory for delayed-acceptance has been considered by \citet{sherlock2015efficiency}, who assess the asymptotic efficiency of DA in random walk Metropolis and pseudo-marginal random walk Metropolis. Their theoretical analysis provides some practical guidelines for tuning DA algorithms but is tailored to MCMC.

Related methods for reducing computation cost for Bayesian models without approximation include early rejection \citep{solonen2012efficient}. Early rejection partitions the posterior into a monotonically decreasing ordering where the cumulative MH ratio can be checked sequentially and rejected as soon as a proposal is deemed infeasible. No approximation to the original MH kernel is made, instead early rejection relies on partial calculation of the posterior density at each iteration. Such a partition has also been used for ABC. In particular, early rejection based on the prior can avoid the expensive simulation required to evaluate the indicator kernel \citep{everitt2017delayed}. Lazy ABC \citep{prangle2016lazy} is another example of work in speeding up Bayesian computation. Lazy ABC uses a random stopping rule to end simulations that are unlikely to result in a feasible parameter, and ensures an unchanged target distribution by reweighting. Like early rejection in ABC, this can be highly effective if the simulations are costly.

As mentioned, delayed-acceptance has been used in sequential Monte Carlo previously by \citet{everitt2017delayed} for ABC where the surrogate model is a cheap, but approximate simulator.  The novel adaptation strategy we describe can also be used in ABC-SMC, but is not limited to this type of Bayesian inference.

The general delayed-acceptance routine extends the standard Metropolis-Hastings, outlined in Section~\ref{sec:smc-intro}, into several steps. The standard MH acceptance probability $\alpha(\vect{\theta}_{*},\vect{\theta})$ is decomposed into two or more distinct acceptance probabilities for delayed-acceptance. For simplicity, we will describe and use only two acceptance probabilities in this paper but the ideas can be applied to several stages of surrogate models. For a given particle $\vect{\theta}_{t}$, at iteration $t$ of the SMC algorithm, the sequential acceptance probabilities in delayed-acceptance are
    \begin{align*}
         \alpha_{s}(\vect{\theta}_{*},\vect{\theta}_{t}) &= \min \left\{1, r_{s}(\vect{\theta}_{*},\vect{\theta}_{t}) \right\}, \text{ for } s = 1,2 \text{ with } \\ r_{1}(\vect{\theta}_{*},\vect{\theta}_{t}) &= \frac{\tilde{p}_{t}(\vect{\theta}_{*}|\vect{y}) q_{\vect{\phi}}(\vect{\theta}_{t} |\vect{\theta}_{*}) }{\tilde{p}_{t}(\vect{\theta}_{t}|\vect{y}) q_{\vect{\phi}}(\vect{\theta}_{*}|\vect{\theta}_{t}^i)}\\
          r_{2}(\vect{\theta}_{*},\vect{\theta}_{t}) &= \frac{p_{t}(\vect{\theta}_{*}|\vect{y}) \tilde{p}_{t}(\vect{\theta}_{t}|\vect{y}) }{p_{t}(\vect{\theta}_{t}|\vect{y}) \tilde{p}_{t}(\vect{\theta}_{*}|\vect{y})} = \left[ \frac{p(\vect{y}|\vect{\theta}_{*})\tilde{p}(\vect{y}|\vect{\theta}_{t})}{p(\vect{y}|\vect{\theta}_{t})\tilde{p}(\vect{y}|\vect{\theta}_{*})} \right]^{\gamma_{t}}
    \end{align*}

where $\tilde{p}_{t}(\vect{\theta}|\vect{y}) \propto [\tilde{p}(\vect{y}|\vect{\theta})\pi(\vect{\theta})]^{\gamma_t} p_0(\vect{\theta})^{1-\gamma_t}$ is the annealed surrogate posterior. The delayed-acceptance MH step proceeds in two parts. First test the proposal using only the (computationally inexpensive) surrogate likelihood, by provisionally accepting the proposal with probability $\alpha_{1}(\vect{\theta}_{*}, \vect{\theta}_{t})$.
 Second, if accepted by the surrogate likelihood, accept the proposal definitively with probability  $\alpha_{2}(\vect{\theta}_{*},\vect{\theta}_{t})$ which accounts for the discrepancy between the surrogate and full likelihood ensuring overall that the correct MH ratio has been used. Proposals that are rejected during the initial screening by the surrogate likelihood are not evaluated by the full likelihood, thus saving computational time when the surrogate is representative of the full likelihood.

We use a robust delayed-acceptance method proposed by \citet{banterle2019accelerating} who have adjusted the standard delayed-acceptance routine to allow some proposals to bypass the surrogate testing with low probability. This mitigates potential stability issues, for example if the approximate likelihood has too light tails relative to the full likelihood.

In this paper we generalise the method for tuning kernel parameters and iteration number put forth by \citet{salomone2018unbiased} as accounting for computation time becomes a more pressing issue with expensive likelihood functions. The framework we propose for tuning kernel parameters and choosing the number of cycles for each SMC mutation step is detailed in the next two sections, for both the MH and delayed-acceptance kernels.

\section{SMC for expensive likelihoods}
\label{sec:tuning-mh-smc}

Our research is motivated by seeking efficient use of adaptive delayed-acceptance algorithm within SMC. In this section, we first develop a novel framework for optimising the number of mutations cycles used in a single SMC iteration and demonstrate how this framework encompasses existing approaches in the literature.  We then utilise and extend the framework to optimise delayed-acceptance within SMC in Section~\ref{sec:tuning-da-mh}.

The mutation step in SMC algorithms requires careful consideration when the likelihood is expensive. On the one hand, the diversification of particles is crucial in maintaining a representative and non-degenerate particle population, but this comes at the computational cost of evaluating the expensive likelihood frequently. In particular, when using delayed-acceptance we would like to lower the first stage acceptance rate whilst increasing the number of iterations to significantly reduce the number of expensive likelihood evaluations.

Several adaptive SMC (and MCMC) algorithms exist that aim to ensure sufficient diversification, but these are yet to explicitly address the issue of the associated computational cost. Our major contribution is to frame the computational cost in SMC with and without delayed-acceptance as an optimisation problem. To begin let $C(k, \vect{\phi})$ be the cost of~$k$ cycles of the MCMC kernel with tuning parameters $\vect{\phi}$ belonging to the set $\Phi$. The tuning parameters considered for the optimisation may be all of the available parameters for the proposal, or a subset of these parameters. This is the case for stage~(d2) of Algorithm~\ref{alg:SMC-alg} as parameters chosen during stage~(d1) remain fixed.

We wish to minimise this cost whilst maintaining some particle diversification condition, $D(k, \vect{\phi}) \geq d$. Under such conditions, an appropriate formulation of the optimisation is
\begin{align}
\argmin_{(k,\vect{\phi}) \in \mathsf{D} }~ C(k, \vect{\phi}) \quad \text{where} \quad \mathsf{D} = \{(k,\vect{\phi}) \in \mathbb{Z}^{+} \times \Phi: D(k,\vect{\phi}) \geq d \}
\label{eq:generalopt}
\end{align}
for some minimum threshold $d$. For example we expect an appropriate, but approximate, cost function for an MH kernel to be
\begin{equation}
  C(k, \vect{\phi}) = k \times L_{F} \label{eq:mh-cost}
\end{equation}
where $L_{F}$ is the cost of evaluating the likelihood, and $k$ is the number of cycles the MCMC kernel iterates for. The feasible set, $\mathsf{D}$, specifies combinations of $k$ and $\vect{\phi}$ which will result in sufficient particle diversification. In the discussion that follows we suppress the notation that $(k,\vect{\phi}) \in \mathbb{Z}^{+} \times \Phi$ when defining feasible sets. We also note if some elements of $\vect{\phi}$ are fixed then $\Phi$ should be replaced by $\Phi_{F} = \{\vect{\phi} \in \Phi : \phi_{i} = \phi_{i}^{\ast} ~\text{for}~i\in F\}$, where $F$ is a set of indices denoting the restricted tuning parameters, and $\phi_{i}^{\ast}$ are the corresponding fixed values. This will be the case when some tuning parameters are determined prior to the optimisation considered in \eqref{eq:generalopt} or chosen by the user.

To make the optimisation tractable in practice we will perform a pilot run of the MCMC kernel over a grid of tuning parameters for~$\vect{\phi}$. This equates to choosing $\Phi$ to be such a grid. The pilot run will assist in estimating the quantities required to perform the optimisation. This is not the only strategy possible of course, but we have found it to perform well. The general optimisation problems described will be transferable to other strategies one might adopt. As mentioned, some elements of~$\vect{\phi}$ may already be fixed from the tuning process in stage (d1) of Algorithm~\ref{alg:SMC-alg}. For example, the proposal variance may be calculated from the weighted particles and fixed at this value.  We defer the discussion of step (d1) until Section~\ref{sec:cal-approx}. The current section is concerned with the optimisation in step (d2).

To develop specific, but potentially approximate, solutions of \eqref{eq:generalopt} we first must consider the feasible sets imposed by an appropriate diversification criterion. In Section~\ref{sec:jdd-std} we study our main criterion based on the expected squared jumping distance (ESJD). We also provide details for an alternative criterion using the MH acceptance probability to ensure diversification. This alternative may be helpful when a notion of distance is not easily defined for the parameter space, but its discussion is deferred to Appendix~\ref{sec:mad} for brevity. Both criteria relate to existing practices in the SMC literature which we address in their respective sections.

In what follows we consider the process of tuning the mutation step at a given iteration of the SMC sampler. We assume that one evaluation of the full likelihood costs $L_{F}$ units, and are concerned with using $k$ cycles of a parameterised MCMC kernel as our mutation step.

\subsection{Jumping distance diversification}\label{sec:jdd-std}

As with all Monte Carlo samplers, it is difficult to optimise the variance of estimators in real time. For SMC, variance is introduced when the set of particles degenerate, and we can avoid high variance by mutating particles to ensure diversity. In this sense, the ESJD is a good candidate measure for particle diversification in SMC as it balances the trade-off between decreasing acceptance rates and increasing jump sizes (and vice versa) under different tuning parameters in the MH proposal. This is somewhat related to the motivation for ESJD in MCMC, in that \citet{pasarica2010adaptively} used ESJD due to the equivalence between maximising the ESJD and minimising a Markov chain's first-order autocorrelation.

The ESJD criterion has been applied to SMC by \citet{fearnhead2013adaptive} and \citet{salomone2018unbiased}. In particular, \citet{fearnhead2013adaptive} adaptively tuned the kernel by drawing parameters from a set that was reweighted and mutated based on their ESJD performance in previous iterations much like SMC itself. \citet{salomone2018unbiased}, on the other hand, found good performance by allocating a tuning parameter from a candidate set to each particle during a pilot run of the mutation step, and selecting the tuning parameter with the highest median ESJD. This parameter is used in subsequent mutation steps (within the same SMC iteration) until a threshold for the total median ESJD is met. This second criterion is the one we choose to generalise as it is more amenable to the calculations required to apply the optimisation in \eqref{eq:generalopt} to delayed-acceptance in SMC.

We start by defining the conditional ESJD, which will be used as the starting point for formalising the criterion in \citet{salomone2018unbiased}. For the $s$th mutation cycle on a given particle, define the conditional ESJD as the conditional expectation
\begin{align}
J(\vect{\theta}_{s-1}, \vect{\theta}_{*}) = \E \left( \Vert \vect{\theta}_{s} -  \vect{\theta}_{s-1}\Vert_{\Sigma}^2~\vert~\vect{\theta}_{s-1}, \vect{\theta}_{*}\right)
\end{align}
where $\vect{\theta}_{s-1}$ is the current position of the particle, $\vect{\theta}_{*}$ is the proposed move for the particle, and $\Vert\cdot\Vert_{\Sigma}^2$ denotes the squared Mahalanobis distance with covariance matrix $\Sigma$.  For an MH kernel, the conditional ESJD can be written as
	\begin{align}
	J(\vect{\theta}_{s-1}, \vect{\theta}_{*}) = \Vert \vect{\theta}_{*} -  \vect{\theta}_{s-1}\Vert_{\Sigma}^2 \alpha(\vect{\theta}_{s-1},\vect{\theta}_{*}), \label{eq:c-esjd}
	\end{align}
where $\alpha(\vect{\theta}_{s-1},\vect{\theta}_{*})$ is the MH acceptance probability of moving from $\vect{\theta}_{s-1}$ to $\vect{\theta}_{*}$, and $s = 1, 2, \ldots, k$ is the current cycle of the mutation kernel. The ESJD of \citet{pasarica2010adaptively} is found by taking the expectation with respect to the conditional values in \eqref{eq:c-esjd} with distributions
\begin{align}
  \vect{\theta}_{s-1} \sim p_{t}(\vect{\theta}\vert\vect{y}) \quad \text{and} \quad  (\vect{\theta}_{*}~\vert~\vect{\theta}_{s-1}) \sim q_{\vect{\phi}}(\vect{\theta}_{s-1}, \cdot) \label{eq:esjd-cond-dist}
\end{align}
where $q_{\vect{\phi}}$ is the proposal distribution of the MH kernel. We will denote the idealised ESJD random variable, or jumping distance, generated during iteration $s$ of the SMC algorithm as
\begin{align*}
&J_{s}(\vect{\phi}) = \Vert \vect{\theta}_{*} -  \vect{\theta}_{s-1}\Vert_{\Sigma}^2 \alpha(\vect{\theta}_{s-1},\vect{\theta}_{*})
\end{align*}
where $\vect{\theta}_{*}$ and $\vect{\theta}_{s-1}$ are approximately distributed according to \eqref{eq:esjd-cond-dist},  due to a finite number of particles.

The first jumping distance-based diversification criterion we consider is the feasible set $\mathsf{D}_{p}$ defined as
\begin{align}
 \mathsf{D}_{p} = \left\{ (k, \vect{\phi}): P(k, \vect{\phi}) \geq p_{\min} \right\} \quad\text{  where  }\quad P(k, \vect{\phi}) = \Pr\left( \sum_{s=1}^{k} J_{s}(\vect{\phi}) \geq d \right)
        \label{eq:div-jdd-gen}
\end{align}
for some quantile $d > 0$.

Unfortunately, using the general jumping distance criterion in \eqref{eq:div-jdd-gen} to select the tuning parameters with a pilot MCMC run is challenging since it depends on $k$ cycles of the mutation kernel. To simplify, we will first limit our focus to the median, but note our arguments can apply to any chosen quantile. Let the feasible set $\mathsf{D}_{m}$ be
\begin{align}
 \mathsf{D}_{m} = \left\{ (k, \vect{\phi}): D(k, \vect{\phi}) \geq d \right\} \quad\text{  where  }\quad D(k, \vect{\phi}) = \med\left\{\sum_{s=1}^{k} J_{s}(\vect{\phi}) \right\}
          \label{eq:div-jdd-med}
\end{align}
and note that $\mathsf{D}_m = \mathsf{D}_p$ when $p_{\mathrm{min}} = 0.5$.
This criterion is still relatively difficult to optimise, so we use an upper bound found by Jensen's inequality for multivariate medians  \citep[][Theorem 5.2]{merkle2010jensen} to develop an approximate optimisation problem. The new, and tractable, optimisation problem replaces $D(k, \vect{\phi})$ in \eqref{eq:div-jdd-med} with
\begin{align}
\tilde{D}(k, \vect{\phi}) = k \times \med\left\{ J_{1}(\vect{\phi}) \right\}
\label{eq:relax-jdd-med}
\end{align}
where $J_{1}(\vect{\phi})$ is the first jumping distance with respect to the tuning parameter(s) $\vect{\phi}$. In practice, each $J_{1}(\vect{\phi})$ is estimated using a pilot run of the mutation step. The approximation can be thought of as assuming the first jumping distance is representative of the jumping distances for subsequent steps. Under this criterion, a diversification threshold of $d$ is achieved when the median of the jumping distance from a single cycle is greater than $d/k$, the average distance per total number of cycles.

Using $\tilde{D}(k, \vect{\phi})$, rather than the original $D(k, \vect{\phi})$, imposes a stronger condition on the diversification requirement during the mutation step, as outlined in Proposition~\ref{th:esjd-jen}.

\begin{prop}\label{th:esjd-jen}
	Consider the feasible sets $\mathsf{D}_m$ and $\tilde{\mathsf{D}}_m$, as
  \begin{align*}
  \mathsf{D}_m &= \left\{ (k, \vect{\phi}): D(k, \vect{\phi}) \geq d \right\} \quad\text{  where  }\quad D(k,\vect{\phi}) = \med\left\{\sum_{s=1}^{k} J_{s}(\vect{\phi}) \right\} \\
  \tilde{\mathsf{D}}_m &= \left\{ (k, \vect{\phi}): \tilde{D}(k, \vect{\phi}) \geq d \right\} \quad\text{  where  }\quad \tilde{D}(k, \vect{\phi}) = k \times \med\left\{ J_{1}(\vect{\phi}) \right\}
  \end{align*}
   For a given $\vect{\phi}$, if $J_{s}(\vect{\phi})$, $s = 1, 2,\ldots,k$, are iid then $\tilde{\mathsf{D}}_m \subseteq \mathsf{D}_m$.
\end{prop}
Imposing the feasible set $\tilde{\mathsf{D}}_m$ on the optimisation problem \eqref{eq:generalopt} ensures the weaker condition of the feasible set $\mathsf{D}_m$ also holds. This motivates the use of $\tilde{\mathsf{D}}_m$ as an approximation of the intended optimisation. The proof of Proposition~\ref{th:esjd-jen} is in Appendix~\ref{pr:esjd-jen}.

With the approximate diversification criterion, $\tilde{\mathsf{D}}_m$ and a MH cost objective function, we can simplify the general optimisation \eqref{eq:generalopt} to coincide with the rule given in \citet{salomone2018unbiased}.
\begin{prop}\label{th:esjd-mh}
	Assume the cost function is $C(k, \vect{\phi}) = k \times L_{F}$, approximating the cost of a standard Metropolis-Hastings step, and diversification criterion is imposed by the feasible set $\tilde{\mathsf{D}}_m$. The solution to \eqref{eq:generalopt} will be equivalent to
	\begin{align}
    \argmax_{\vect{\phi} \in \Phi}~~  \med\left\{ J_{1}(\vect{\phi}) \right\}.
	\end{align}
\end{prop}
A proof is in Appendix~\ref{pr:esjd-mh}. Proposition~\ref{th:esjd-mh} provides further justification for the median ESJD tuning rule \citep{salomone2018unbiased}, the authors having found it useful for choosing a minimum number of mutation cycles to avoid degenerate particle populations.

To apply this principle in practice we can use a pilot run of the mutation step choosing $\Phi$ as a discrete grid in order to make this problem tractable. Our approach is to randomly assign each particle a value from a relatively small pool of candidates, generally a grid of reasonable values. For example, when $\vect{\phi} = [h]$ --- the step size in a MVN proposal --- we could take~$\Phi = \{0.2,0.4,0.6,0.8\}$ and assign these values to each particle using a random partition (of equal size) of the particles. The median jumping distance, $m_{\vect{\phi}} = \med\left\{ J_{1}(\vect{\phi}) \right\}$ for each $\vect{\phi} \in \Phi$, can then be approximated by
\begin{align*}
    \widehat{m_{\vect{\phi}}} =
		\med\left\{ J^{i}_{1} \right\}_{i \in S_{\vect{\phi}}}
\end{align*}
where $J^{i}_{1}$ is the jumping distance from the $i$th particle in the pilot run, and $\{ S_{\vect{\phi}} : \vect{\phi} \in \Phi \}$ is a partition where each $ S_{\vect{\phi}}$ describes the allocation of particles to each parameter value, $\vect{\phi}$.

The main benefit of using a grid for the optimisation is to ensure that the procedure is simple and unburdensome computationally. There are other possibilities however, which may be more appropriate in higher dimensions of $\vect{\phi}$.  For example, if assigning each particle a unique value of $\vect{\phi}$, running a non-linear regression to generate an estimate of the median jumping distance may be feasible.

\section{Delayed-acceptance SMC}
\label{sec:tuning-da-mh}

So far we have set up the computational optimisation problem we wish to solve, and shown how it applies to jumping distance diversification in Section~\ref{sec:jdd-std}. The solutions we derive for the optimal tuning parameters correspond to mutations using cycles of MH kernels in SMC. We now extend the optimisation framework to cycles of the delayed-acceptance kernel.

Under two-stage delayed-acceptance the optimisation now has the objective function
\begin{align}
    C(k,\vect{\phi}) = k(L_{S} +  \alpha_{1}(\vect{\phi})L_{F}) \label{eq:da-cost}
\end{align}
which can be interpreted as $k$ MCMC iterations, each with cost $L_{S}$ for evaluating the approximate likelihood and additional cost $L_{F}$ for evaluating the full likelihood which occurs with average probability $\alpha_{1}(\vect{\phi})$, and is dependent on the tuning parameters of the proposal distribution. Combined with the chosen diversification criterion, this optimisation is somewhat more involved. Our approach is to (again) rely on taking the possible tuning parameter set, $\Phi$, to be a small finite set of values. This set should be much smaller than the total number of particles so that the pilot run can effectively estimate the parameters required to solve the (approximate) optimisation.


For delayed-acceptance with an adaptive Gaussian proposal distribution the tuning parameter $\vect{\phi}$ is the step size $h$. A pilot run of the mutation step is performed where the $i$th particle is randomly allocated to group $S_{\vect{\phi}}$ associated with a particular grid point, $\vect{\phi} \in \Phi$, whilst ensuring evenly sized groups.

The first stage acceptance probability can be estimated for each $\vect{\phi} \in \Phi$ by averaging within the groups to yield $\widehat{\alpha}_{1}(\vect{\phi})$, whilst the average likelihood costs $L_{F}$ and $L_{S}$ can be estimated using all of the pilot runs. These estimates allow us to approximate the first component of the optimisation, the computational effort $C(k, \vect{\phi})$. In order to ensure the jumping distance threshold is met, we need to estimate the overall acceptance probability, and provide a means to approximate the total (additive) jumping distance after $k$ mutation cycles -- based only on one pilot run. The general strategy is to take $\Phi$ as a discrete grid, estimate the minimum number of cycles required to reach the diversification threshold for each $\vect{\phi} \in \Phi$, then find the $\vect{\phi}$ with minimal cost.

Once a pair of tuning parameters and required iterations has been found, $(\vect{\phi}^{*},k_{\vect{\phi}}^{*})$, one can either iterate the mutation step $k_{\vect{\phi}}^{*}$ times ($k_{\vect{\phi}}^{*}-1$ is also an option), or monitor the empirical diversification threshold, $D\left(\{J^{i}\}_{i=1}^{N}\right)$, after each mutation step until it has been met. We choose the latter for our simulation study in Section~\ref{sec:sim-ex} and example using the Whittle likelihood in Section~\ref{sec:whittle}. This adaptive monitoring is described in Algorithm~\ref{alg:pilotmutation} along with details for incorporating a pilot run into the mutation component of SMC to optimise the tuning parameters.

To include such adaptivity in SMC we replace steps (d2) and (e) of Algorithm~\ref{alg:SMC-alg} with Algorithm~\ref{alg:pilotmutation}. Note that the empirical diversification criterion, $D\left(\{J^{i}\}_{i=1}^{N}\right)$, is the median of the accumulated squared jumping distances across the particles when using ESJD diversification.

\begin{algorithm}
    \caption{SMC mutation with pilot step for tuning parameters and ESJD diversification criterion monitoring \label{alg:pilotmutation}}
    \textbf{Input:} Current particle set, $\{\vect{\theta}_{t}^i\}_{i=1}^{N}$; Family of MCMC kernels, $K_{\vect{\phi}}$; Kernel tuning set,~$\Phi$; Diversification criterion threshold, $d$.

        \begin{enumerate}
            \item[(d2)] Pilot mutation step, set $s=0$, then
            \begin{enumerate}[(i)]
                \item Allocate particles by partition $\bigcup_{\vect{\phi} \in \Phi}S_{\vect{\phi}} = \{1,\ldots,N\}$
                \item Set $\vect{\phi}^{i} = \vect{\phi}$ for $i \in S_{\vect{\phi}}$, for each $\vect{\phi} \in \Phi$
                \item Mutate particles$^{1}$ $\check{\vect{\theta}}_{0}^i \sim K_{\vect{\phi}^{i}} (\vect{\theta}_{t}^i, \cdot)$ and store proposals $\check{\vect{\theta}}_{*}^{i}$ for $i\in \{1,\ldots,N\}$
                \item Calculate ESJD value $J^{i}_{0} = J(\vect{\theta}_{t}^{i}, \check{\vect{\theta}}_{*}^{i})$ as in \eqref{eq:c-esjd} with $\hat{\Sigma}$ from population of particles
                \item Calculate optimal tuning parameter $\vect{\phi}^{\ast}$ (\textit{see Algorithm~\ref{alg:min-k}})
            \end{enumerate}
            \item[(e)] While $\med\left(\{J^{i}_{s}\}_{i=1}^{N}\right) < d$, increment $s$, then iterate through
            \begin{enumerate}[(i)]
                \item Mutate particles$^{1}$ $\check{\vect{\theta}}_{s}^i \sim K_{\vect{\phi}^{\ast}}(\check{\vect{\theta}}_{s-1}^i, \cdot)$ and store proposals $\check{\vect{\theta}}_{*}^{i}$ for $i\in \{1,\ldots,N\}$
                \item Update diversification value $J^{i}_{s} = J^{i}_{s-1} + J(\check{\vect{\theta}}_{s-1}^i, \check{\vect{\theta}}_{*}^{i})$ for $i\in \{1,\ldots,N\}$
            \end{enumerate}
        \end{enumerate}
    \textbf{Output:} Updated particle values $\{\check{\vect{\theta}}_{s}^i \}_{i=1}^{N}$ \vspace{0.2cm}\\
    $^{1}$Some additional information from the mutation step is required for diversification criterion calculation and tuning parameter optimisation, e.g. proposal values and acceptance probabilities.
\end{algorithm}

\subsection{Estimating the optimal kernel parameters}\label{sec:jdd-da}

The main hurdle in choosing the optimal kernel parameters with cost function \eqref{eq:da-cost} and finite discrete set for $\Phi$ is determining the minimum number of MCMC cycles required to satisfy the diversification criterion. This section details several possible methods for this important problem, followed by Algorithm~\ref{alg:min-k} which describes the overall process of choosing the parameters.

To estimate the number of cycles $k$ required for each element $\vect{\phi} \in \Phi$ to achieve the required jumping distance threshold in \eqref{eq:relax-jdd-med} we can use a pilot run of the mutation step. After such a pilot run we have a set of realisations of the ESJD for each tuning parameter, that is, we have observed values for $J_{1}(\vect{\phi})$. Fitting an amenable model to these realisations will allow us to estimate how many cycles are required for diversification under each parameter. We propose three such methods for modelling the expected squared jumping distance used to estimate the required number of cycles.

\subsubsection{Median method}\label{sec:esjd-median-method}

The first model uses the same principles as the median method described in Section~\ref{sec:jdd-std}, whereby the minimum number of cycles $k$ is approximated by
\begin{equation}
  k^{*}_{\vect{\phi}} = \min\left\{k \in \mathbb{Z}_{+}: k \times \med\{J_{1}(\vect{\phi})\}\geq d \right\}.
\end{equation}

In practice the number of cycles is then chosen by $k^{*}_{\vect{\phi}} = \lceil d / \med\{J_{1}(\vect{\phi})^{i}\}_{i \in S_{\vect{\phi}}} \rceil$, where $J_{1}(\vect{\phi})^{i}$ is the realisation of the ESJD for the $i$th particle. We refer to this process as the median method.

\subsubsection{Gamma method}\label{sec:esjd-gamma-method}

We may also model the jumping distances parametrically, which we describe as follows. The ESJD are positive random variables generated from the normalised proposal distribution distance multiplied by the acceptance probability. In the case of a multivariate normal proposal, the use of the squared Mahalanobis distance results in the proposed jumps having a Chi-squared distribution with degrees of freedom $p$. This motivates one choice of model for the \textit{expected} squared jumping distance -- the gamma distribution. It has the desirable properties of having positive support and the sum of gamma random variables is also gamma, providing a means to estimate how many iterations are required to achieve the minimum diversification threshold.

Under the gamma model, for each group of jumping distances, $\{J_{1}(\vect{\phi})^{i}\}_{i \in S_{\vect{\phi}}}$, indexed by $\vect{\phi}$ we fit a gamma distribution to obtain the parameter estimates for $a_{\vect{\phi}}$ and $b_{\vect{\phi}}$, the shape and rate respectively. Based on these parameters, the total ESJD of $k$ cycles is approximated by
\begin{equation}
\sum_{s=1}^{k} J_{s}(\vect{\phi}) \sim \dGam(k\widehat{a_{\vect{\phi}}},\widehat{b_{\vect{\phi}}}) \label{eq:gamma-model-k}
\end{equation}
using the additive property of the gamma distribution. We can then calculate the minimum iterations for each parameter $\vect{\phi}$,
\begin{equation}
  k^{*}_{\vect{\phi}} = \min \left\{ k \in \mathbb{Z}_{+}: \Pr\left( \sum_{s=1}^{k} J_{s}(\vect{\phi})  \geq d \right)  \geq p_{\mathrm{min}} \right\} \label{eq:min-k}
\end{equation}
by iterating through $k = 1, 2, \ldots$ using the complementary cumulative distribution function (CCDF) of the gamma distribution in \eqref{eq:gamma-model-k} and noting the monotonicity in $k$ of the required probability. If $\widehat{a_{\vect{\phi}}}$ is small, it may be pragmatic to do a line search rather than iterate through each $k = 1$ then $k = 2$ et cetera.

\subsubsection{Bootstrap method}\label{sec:esjd-boot-method}

We also suggest a non-parametric alternative, which we refer to as the bootstrap method. Under this method we assume each $J_{s}(\vect{\phi})$ for $s = 1, 2, \ldots$ is drawn from a discrete distribution with values from the pilot run $\{J_{1}(\vect{\phi})^{i}\}_{i \in S_{\vect{\phi}}}$ for $\vect\phi \in \Phi$. If we assign equal probability to these draws then we can bootstrap the minimum $k^{*}_{\vect{\phi}}$ required. For each $\vect\phi$ we continue drawing new jumping distances until the probability threshold in \eqref{eq:min-k} is met. This approximates the required number of mutation cycles non-parametrically.

The processes for calculating $k^{*}_{\vect{\phi}}$ and choosing the tuning parameter(s) is described in Algorithm~\ref{alg:min-k} for both the gamma and bootstrap methods of the tuning parameters. Replacing steps 1(a)-(b) of Algorithm~\ref{alg:min-k} with $k_{g}= \lceil d / \med\{J_{1}(\vect{\phi}_{g})\} \rceil$ describes the algorithm for the median method. Algorithm~\ref{alg:min-k} is used as step (d2)(v) in Algorithm~\ref{alg:pilotmutation}.

\begin{algorithm}
  \caption{Tuning parameter $\vect\phi$ selection from pilot mutation run \label{alg:min-k}}
	\textbf{Input:} Tuning parameter set, $\Phi = \{\vect\phi_{g}\}_{g=1}^{G}$; Allocation of tuning parameters to particles, $\{S_{\vect{\phi}}\}_{\vect{\phi} \in \Phi}$; Pilot run ESJDs, $\{J^{i}\}_{i=1}^{N}$; Average first stage acceptance probabilities, $\{\widehat{\alpha}_{1}(\vect{\phi}_{g})\}_{g=1}^{G}$; Approximate computation cost of surrogate and full likelihoods, $L_{S}$ and $L_{F}$; Minimum probability quantile, $p_{\min}$.
	\begin{enumerate}
    \item For $g \in \{1,\ldots,G\}$ iterate through
    \begin{enumerate}[(a)]
      \item Set $k_{g} = 0$ and $\hat{p} = 0$
      \item While $\hat{p} \leq p_{\min}$, increment $k_g$ then update the quantile estimate:\\
      $\hat{p}  = \Pr\left( \sum_{s=1}^{k_{g}} J_{s}(\vect{\phi}_{g})  \geq d \right)$ using gamma or bootstrap method (\textit{Sections~\ref{sec:esjd-gamma-method} and~\ref{sec:esjd-boot-method}})
      \item Calculate $C_{g} = C(k_{g},\vect{\phi}_{g}) = k_{g}(L_{S} +  \widehat{\alpha}_{1}(\vect{\phi}_{g})L_{F})$
    \end{enumerate}
    \item Calculate index of best tuning parameter $g^{*} = \argmin_{g \in G} C_{g}$
  \end{enumerate}
	\textbf{Return} Selected tuning parameter $\vect\phi_{g^{*}}$
\end{algorithm}

\subsubsection{Estimating overall probability of acceptance}\label{sec:esjd-est-acc}

One further consideration for determining the tuning parameters with a pilot run when using delayed-acceptance, rather than standard Metropolis-Hastings, is required. In the case of delayed-acceptance, the overall acceptance probability is the product of the two acceptance probabilities (see Section~\ref{sec:da-mh-intro}), that is
    \begin{align}
         \alpha(\vect{\theta}_{*},\vect{\theta}_{s}) = \alpha_{1}(\vect{\theta}_{*},\vect{\theta}_{s})\alpha_{2}(\vect{\theta}_{*},\vect{\theta}_{s}).
    \end{align}
However $\alpha_{s}(\vect{\theta}_{*},\vect{\theta}_{s})$ is only calculated for the proposals that pass the surrogate model, hence we cannot calculate $ \alpha(\vect{\theta}_{*},\vect{\theta}_{s})$ exactly for these particles without wasting computation time. As such, we run a linear regression with response $\log r(\vect{\theta}_{*}, \vect{\theta}_t)$, the full log ratio used in standard MH, and explanatory variable $\log r_{1}(\vect{\theta}_{*}, \vect{\theta}_t)$ the surrogate log-ratio for the acceptance rate. We also use the step size of the mutation kernel as an additional explanatory variable, which assists when the first stage acceptance rate is small.  The overall acceptance rate, $\widehat{\log r} = \widehat{\log r(\vect{\theta}_{*}, \vect{\theta}_t)}$, is then predicted for cases where the proposal was rejected in using the surrogate likelihood. We convert this to a probability using the standard MH formula, $\min\{\exp(\widehat{\log r}),1\}$.

Simpler strategies are possible, but were found to be ineffective. For example, using $\alpha(\vect{\theta}_{*},\vect{\theta}_{s}) = \alpha_{1}(\vect{\theta}_{*},\vect{\theta}_{s})\bar{\alpha}_2$ if the proposal was rejected at the surrogate stage when calculating the ESJD, where $\bar{\alpha}_2$ is the mean of the second stage acceptance probabilities that were calculated.

\subsection{Calibrating surrogate likelihoods}
\label{sec:cal-approx}

The surrogate likelihood can be a biased approximation of the full likelihood which impacts the effectiveness of the delayed-acceptance method. Within SMC, the previous evaluations of the full likelihood contain valuable information which we can use to tune the surrogate likelihood. We propose a generic method for calibrating the surrogate likelihood to better match the full likelihood during the delayed-acceptance mutation step of our SMC algorithm. The method is constructed to avoid evaluating the costly full likelihood by relying on the history of the particles, and so that no user-chosen tuning parameters are required. We begin by defining the general transformation considered with the corresponding optimisation problem, followed by the particular implementation we chose to explore in this paper.

The general transformation consists of two parts. The first anneals components of the surrogate likelihood as follows. Suppose the surrogate likelihood, $\tilde{L}(\vect{y}~\vert~\vect{\theta})$, can be decomposed into the product $\tilde{L}(\vect{y}~\vert~\vect{\theta})  = \prod_{j=1}^{q}\tilde{L}_{j}(\vect{y}~\vert~\vect{\theta})$ then define the weighted-annealing transformation of the surrogate as
\begin{align*}\label{eq:ann-surr}
\tilde{L}^{\vect{\zeta}}(\vect{y}~\vert~\vect{\theta})  = \prod_{j=1}^{q}\tilde{L}_{j}(\vect{y}~\vert~\vect{\theta})^{\zeta_{j}}.
\end{align*}
In some cases the product decomposition defined by components $\tilde{L}_{j}(\vect{y}~\vert~\vect{\theta})$ will correspond to each datum, i.e.~$\tilde{L}_{j}(\vect{y}~\vert~\vect{\theta}) = p_{j}(y_{j}~\vert~\vect{\theta})$ where $p_{j}$ is the probability density function for datum $y_{j}$. This need not be the case, and if the surrogate likelihood is not decomposable we can take $q=1$ or redefine the surrogate likelihood with additional components if appropriate.  A special case of this transformation, taking $q = 1$ or $\zeta_{i} = \zeta$, corresponds to power-likelihood annealing commonly used in SMC.

The second transformation on the surrogate likelihood is a bijection of the model parameters, $T_{\vect{\xi}}$, with tuning parameters~$\vect{\xi}$. Combining the annealing weights and parameter transformation, the overall parameterised surrogate likelihood is $\tilde{L}^{\vect{\zeta}}(\vect{y}~\vert~T_{\vect{\xi}}(\vect{\theta}))$.

Prior to the mutation step in each iteration of SMC we would like to minimise a distance (or discrepancy) to find~$(\vect{\xi}^{\star},\vect{\zeta}^{\star})$, the optimal transformation, by solving
\begin{equation}\label{eq:opt-calibration-gen}
  (\vect{\xi}^{\star},\vect{\zeta}^{\star}) = \argmin_{\vect{\xi},\vect{\zeta}} \sum_{\vect{\theta} \in H}d\left[ L(\vect{y}~\vert~\vect{\theta}), \tilde{L}^{\vect{\zeta}}(\vect{y}~\vert~T_{\vect{\xi}}(\vect{\theta}))\right]
\end{equation}
where $d(x,y)$ is a measure of discrepancy, $L(\vect{y}~\vert~\vect{\theta})$ is the full likelihood, and $H$ is the set of particle locations to average across. In some circumstances, it may also be appropriate to add a penalty term to this optimisation problem. To reduce the cost of calibrating the surrogate likelihood in this way, $H$ should be a subset of locations where the expensive likelihood, $L(\vect{y}~\vert~\vect{\theta})$, has already been evaluated. For simplicity, we choose $H$ to be the set of locations from the current particle set in the SMC algorithm.

Our examples in Section~\ref{sec:sim-ex} and~\ref{sec:whittle} use the discrepancy measure to be $d(x,y) = (\log x - \log y)^{2}$, corresponding to the sum of square differences of the log-likelihood, with $T_{\vect{\xi}}(\vect{\theta}) = \vect{\theta} - \vect{\xi}$, and the annealing weights, $\vect{\zeta}$ applied to the density of each datum. The transformation $T_{\vect{\xi}}$ serves to correct for bias in the surrogate likelihood, whilst the  $\vect{\zeta}$ act to flatten or steepen the surrogate as needed. This is a relatively simple choice, for which we formulate an approximate optimisation, and serves to illustrate that even simple calibration can be useful for delayed-acceptance methods.

We approximate the solution for \eqref{eq:opt-calibration-gen} by first minimising the discrepancy with respect to $\vect{\xi}$ with $\vect{\zeta} = \vect{1}$, followed by a conditional optimisation with a lasso penalty \citep{tibshirani1996regression} --- with shrinkage towards the unit vector rather than zero. The proposed approximate solution is
\begin{align}
  \vect{\xi}^{\star} &= \argmin_{\vect{\xi},\mu_{1}} \sum_{\vect{\theta} \in H}\left[ \ell(\vect{y}~\vert~\vect{\theta}) - \tilde{\ell}^{\vect{\zeta}}(\vect{y}~\vert~(\vect{\theta}- \vect{\xi})) - \mu_{1}\right]^{2} \text{ with } \vect{\zeta} = \vect{1}, \label{eq:opt-calibration-1}\\
  \vect{\zeta}^{\star} &= \argmin_{\vect{\zeta}, \mu_{2}} \sum_{\vect{\theta} \in H} \left[\ell(\vect{y}~\vert~\vect{\theta}) - \tilde{\ell}^{\vect{\zeta}}(\vect{y}~\vert~(\vect{\theta}- \vect{\xi}^{\star}))  - \mu_{2}\right]^{2} + \Lambda \Vert \vect{\zeta} - \vect{1} \Vert_{1} \label{eq:opt-calibration-2}
\end{align}
where $\ell(\vect{y}~\vert~\vect{\theta})$ is the full log-likelihood, $\tilde{\ell}^{\vect{\zeta}}(\vect{y}~\vert~(\vect{\theta}) = \log \tilde{L}^{\vect{\zeta}}(\vect{y}~\vert~\vect{\theta})$ is the weighted surrogate log-likelihood, whilst $\mu_{1}$ and $\mu_{2}$ are nuisance parameters. Structuring the optimisation as such, \eqref{eq:opt-calibration-1} can be solved using non-linear least squares, and serves to align the surrogate and full likelihood using a translation.
Whilst \eqref{eq:opt-calibration-2} has the form of lasso regression, since
\begin{equation*}
  \tilde{\ell}^{\vect{\zeta}}(\vect{y}~\vert~(\vect{\theta}) = \sum_{j=1}^{q}\zeta_{j} \log \tilde{L}_{j}(\vect{y}~\vert~\vect{\theta}),
\end{equation*}
where $\zeta_{j}$ are the linear coefficients. Shrinkage towards the unit vector can be achieved with a change of variables, $\vect{\zeta}^{\prime} = \vect{\zeta} - \vect{1}$, which transforms the regression to the standard form. We find it convenient to automate the choice of $\Lambda$ using cross-validation of the lasso \citep{friedman2010regularization}. 

The calibration of the surrogate likelihood proceeds prior to (and independently from) Algorithm~\ref{alg:pilotmutation} as it does not require a pilot mutation step to be performed. We describe surrogate likelihood calibration in Algorithm~\ref{alg:tune-sl} along with other tuning that does not require a pilot mutation step. In particular, we include tuning the covariance matrix of the MVN proposal distribution, which is common in SMC. Algorithm~\ref{alg:tune-sl} replaces step (d1) in Algorithm~\ref{alg:SMC-alg}.

\begin{algorithm}
    \caption{Surrogate likelihood calibration and proposal distribution tuning \label{alg:tune-sl}}
    \textbf{Input:} Particle set prior to resampling, $\{\vect{\theta}^{i}_{t-1}\}_{i=1}^{N}$; Weights prior to resampling $\{\check{W}_{t}^{i}\}_{i=1}^{N}$; History (possibly a subset) of particles, $H$; Likelihood values for the history, $\{\ell(\vect{y}~\vert~\vect{\theta})\}_{\vect{\theta} \in H}$.
    \begin{enumerate}[(d1)] 
    \item Initial tuning of mutation kernel $K_{\vect{\phi}}$
            \begin{enumerate}[(i)]
              \item Calculate $\hat{\Sigma}_{t}$ from weighted particles, $\{\vect{\theta}^{i}_{t-1}\}_{i=1}^{N}$ and $\{\check{W}_{t}^{i}\}_{i=1}^{N}$, for MVN proposal
              \item If $K_{\vect{\phi}}$ is a DA kernel: Compute $(\vect{\xi}^{\star},\vect{\zeta}^{\star})$ using \eqref{eq:opt-calibration-1} and \eqref{eq:opt-calibration-2}.
            \end{enumerate}
    \end{enumerate}
            
    \textbf{Output:} Covariance matrix $\hat{\Sigma}_{t}$, optimal surrogate likelihood parameters $\vect{\xi}^{\star},\vect{\zeta}^{\star}$.
    
\end{algorithm}

\subsection{Surrogate First Annealing}
\label{sec:sfa}

The final application of a surrogate likelihood for efficient SMC algorithm is to utilise it in the distribution path of SMC. It is possible to use the surrogate likelihood for this purpose by annealing through an inexpensive sequence of distributions using the surrogate before correcting to the full posterior. Such a distribution path can eliminate low probability regions of the parameter space with little computational cost. We propose surrogate first annealing (SFA) in which the sequence of distributions the particles travel through are determined by
\begin{align}
  \begin{split}
    p_{t}(\vect{\theta}) &= p_{0}(\vect{\theta})^{\max\{1-\gamma_{t},0\}}
    \tilde{p}(\vect{\theta}~\vert~\vect{y})^{\lambda\min\{\gamma_{t},2 - \gamma_{t}\}}
    p(\vect{\theta}~\vert~\vect{y})^{\max\{0, \gamma_{t} - 1\}} \\
    &\text{for}\quad  0 = \gamma_{0} < \cdots < \gamma_{S} = 1 < \gamma_{S+1} < \cdots < \gamma_{T} = 2\\
    &\text{and}\quad 0 < \lambda \leq 1
  \end{split}
\end{align}
where $p_{0}(\vect{\theta})$ is the initial distribution, $\tilde{p}(\vect{\theta}~\vert~\vect{y})$ is the posterior with surrogate likelihood, $p(\vect{\theta}~\vert~\vect{y})$ is the posterior with full likelihood, and $\lambda$ controls the maximum power of the surrogate posterior during the sequence. In practice this amounts to two SMC runs (in line with Algorithm~\ref{alg:SMC-alg}), where the first anneals from the initial distribution to (a power of) the surrogate posterior, and the second SMC algorithm anneals from (a power of) the surrogate posterior to the full posterior. Some representative temperatures, $\gamma_{t}$, and associated distributions along the annealing path are given in Table~\ref{tab:sfa-temp-ex}.

\begin{table}[H]
  \centering
\begin{tabular}{|l|l|}
\hline
$\gamma_{t}$ & $p_{t}(\vect{\theta})$ \\ \hline
0.0 & $p_{0}(\vect{\theta})$ \\
0.5 & $p_{0}(\vect{\theta})^{0.5}\tilde{p}(\vect{\theta}~\vert~\vect{y})^{0.5\lambda}$ \\
1.0 & $\tilde{p}(\vect{\theta}~\vert~\vect{y})^{\lambda}$\\
1.5 & $\tilde{p}(\vect{\theta}~\vert~\vect{y})^{0.5\lambda} p(\vect{\theta}~\vert~\vect{y})^{0.5}$\\
2.0 & $p(\vect{\theta}~\vert~\vect{y})$\\ \hline
\end{tabular}

\caption{Representative temperatures and distributions using surrogate first annealing.} \label{tab:sfa-temp-ex}
\end{table}

As with all SMC and importance sampling algorithms, if the initial distribution does not cover the tails of the target distribution adequately then the sampler can perform poorly. As such, care needs to be taken that the surrogate posterior is a good initial distribution for the full posterior. Therefore, we recommend choosing $\lambda \leq 0.5$, and $\lambda$ should decrease as the number of observations increases. In general, $\lambda$ should be chosen as small as possible but sufficiently large to eliminate low-probability areas of the parameter space and speed up the SMC algorithm.

\section{Simulation study}
\label{sec:sim-ex}

\subsection{Simulation design}
To test the efficacy of the proposed adaptive SMC algorithm, we consider a linear regression where we artificially control the cost of the likelihood evaluations. In particular, we fit the following normal or student-t regression models
\begin{align}
  (y_{i}~\vert~\vect{\beta},\sigma^{2}) &\sim \dNorm(X \vect{\beta}, \sigma^{2}) \label{eq:norm-like}\\
&\text{or} \notag \\
  (y_i~\vert~\nu, \vect{\beta},s) &\sim \mathcal{T}(\nu, X \vect{\beta}, s) \label{eq:t-like}
\end{align}
for $i = 1,2,\ldots,n$, where $\dNorm(\mu, \sigma^{2})$ denotes the normal distribution with mean $\mu$ and variance $\sigma^2$, and $\mathcal{T}(\nu, \mu, s)$ denotes the location-scale student-t distribution with degrees of freedom $\nu$, mean $\mu$ (if $\nu > 1$) and scale $s$. Both models have the following priors on $\vect{\beta}$
\begin{align*}
  \beta_j & \sim \dNorm(0, \tau^{2}).
\end{align*}
for $j = 1, 2, \ldots, p$. For the normal regression we fix the variance $\sigma^{2}= 0.5^2$, for the t-regression we set $\nu = 3$ and $s = 1$. The prior variance is set to $\tau^{2}= 2^2$.

We simulate data for each regression with $n = 100$, and $p = 5$ with respect to the data-generating distribution specified by \eqref{eq:norm-like} or \eqref{eq:t-like}. The true parameter vector for the simulation is $\vect{\beta} = [0, 0.5,-1.5, 1.5, 3]^{\top}$ and the elements of the design matrix $X$ are iid normal random variables with unit variance.

To test delayed-acceptance on these models we set the full likelihood as \eqref{eq:norm-like} or \eqref{eq:t-like} above, but with an artificial time delay of $L_F$ seconds. The computation is not literally delayed, but instead is artificially inflated each time the likelihood is evaluated. The surrogate likelihood used is normal with additional bias and scaling on $\vect{\beta}$, that is
\begin{equation*}
  \tilde{L}(\vect{y}~\vert~\vect{\beta}) = \dNorm(\vect{y};X(a \vect{\beta} + b\vect{1}), 1^2)
\end{equation*}
where $a=e^{0.1}\approx 1.105$ and $b=0.25$. The surrogate likelihood is also given an artificial delay, $L_{S}$, in order to control the computational difference between the likelihoods. We express this as the ratio $\rho = L_{F}/L_{S}$, as the results can be interpreted for any time unit.

Relative to the surrogate likelihood the full likelihoods, the normal and student-t distributions, represent two idealised extremes possible to encounter with delayed-acceptance. In the normal case, the surrogate likelihood can be transformed to exactly match the full likelihood using the calibration method in Section~\ref{sec:cal-approx}, whereas the surrogate likelihood will never be able to replicate the heavy tails in the student-t likelihood.

We ran several adaptive SMC algorithms under various combinations of settings. The \textit{MH}- and \textit{SFA}-SMC algorithms used SMC with a Metropolis-Hastings transition kernel using an adaptive step size, the latter using the surrogate first annealing as described in Section~\ref{sec:sfa}. The \textit{DA}, \textit{DA+T}, \textit{DA+SFA}, and \textit{DA+T+SFA} algorithms use the delayed-acceptance kernel where \textit{+T} indicates surrogate calibration was used as in Section~\ref{sec:cal-approx}, \textit{+SFA} indicates using surrogate first annealing. Each of these algorithms were tested with kernel tuning using the median, gamma, and bootstrap methods described in Section~\ref{sec:jdd-da} using the cost function \eqref{eq:mh-cost} for MH and \eqref{eq:da-cost} for DA.  We also ran two SMC algorithms akin to MH-SMC, but with fixed step size $h$. The step size for these algorithms, \textit{MH (f-)} and \textit{MH (f+)}, were chosen by taking the average optimal step size from the MH-SMC algorithm and selecting the closest smaller (f-) or larger (f+) step size from the set $\Phi$. Note that the performance tables that follow only report the best performing of MH (f-) and MH (f+).

We tested the above SMC algorithms with the following settings. The number of particles $N = 2000$, the initial distribution $p_{0}$ is the prior, and the proposal distribution for the mutation step is an adaptive-variance multivariate normal random walk, i.e. \eqref{eq:mvn-mh-kernel}. The ESS threshold to adaptively select the temperature, $\gamma_{t}$, is $S = N/2$. Each SMC algorithm used stratified sampling with 10 strata to resample the particles.

\begin{figure}
    \centering
    \includegraphics[width = 0.9\textwidth]{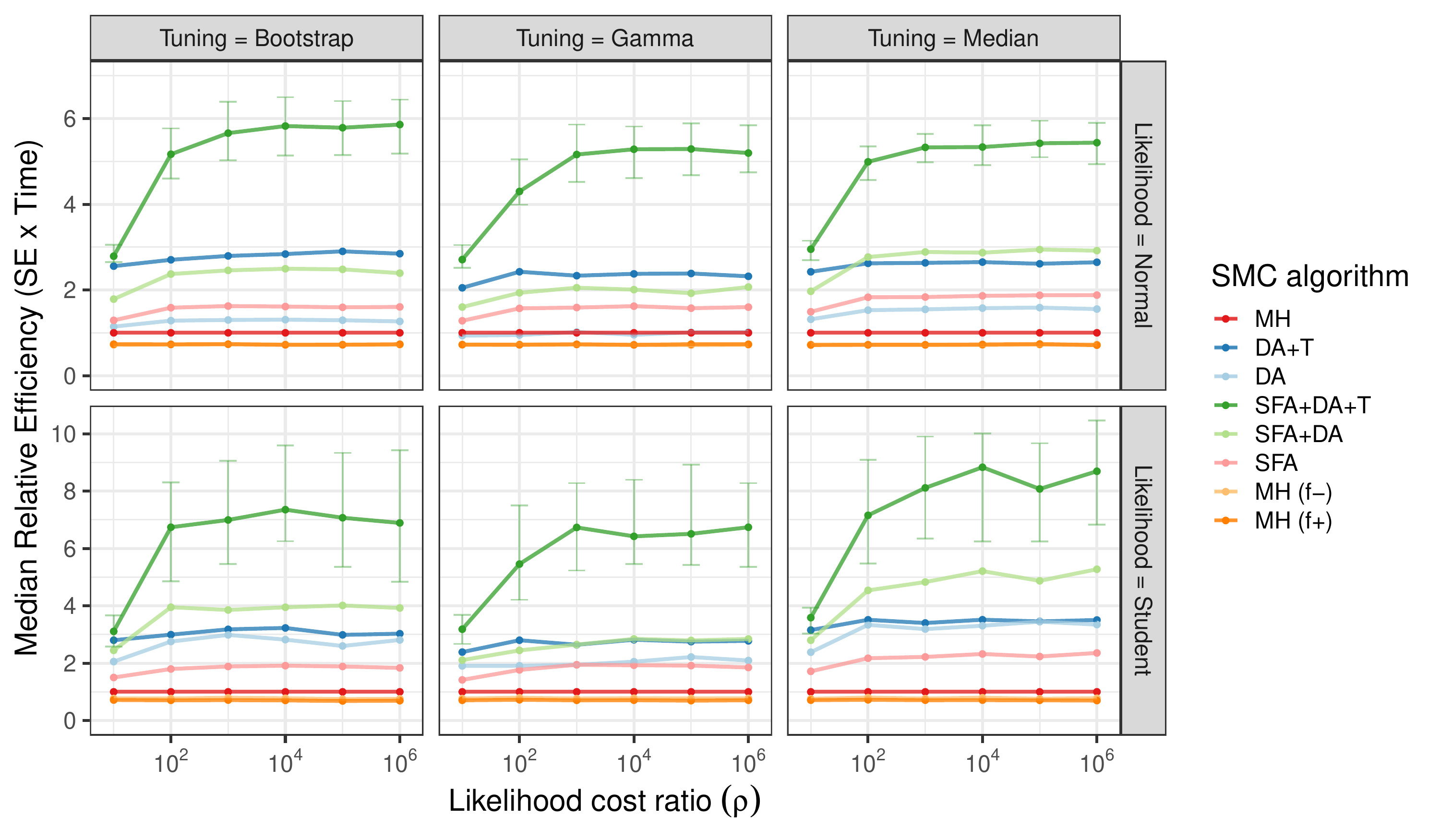}
    \caption{Computational efficiency as measured by squared error $\times$ computation time (SE $\times$ Time) for competing SMC algorithms. Median efficiency is plotted relative to the MH-SMC algorithm (constant red line at 1) which used median tuning across all simulations. The 10th and 90th quantiles are plotted as error bars for the SFA+DA+T algorithm.}
    \label{fig:eff-2000}
\end{figure}

We choose the diversification threshold to be the median ($p_{\mathrm{min}} = 0.5$) of the total ESJD greater than $d \approx 2.34$ chosen such that $\Pr(X > d) = 0.8$, where $X \sim \chi^{2}(5)$, a Chi-square distribution with 5 degrees of freedom. The choice of $d$ is motivated by the following idea. Conditional on the current location of the particle, $\vect{\theta}_{s-1}$, the $\Vert \vect{\theta}_{*} -  \vect{\theta}_{s-1}\Vert_{\Sigma}^2$ term in the ESJD definition \eqref{eq:c-esjd} has a $\chi^{2}(5)$ distribution  (the number of parameters). As such, requiring $\Pr(X > d) = 0.8$ ensures that after $k$ mutation cycles with any level of acceptance, the empirical distribution of the ESJD exceeds the 80\% quantile of the distribution assuming all proposals are accepted from one cycle.

The number of mutation cycles is optimised with respect to the step-scale, $h$, for the MVN random walk, as described in Sections~\ref{sec:tuning-mh-smc} and~\ref{sec:tuning-da-mh}. Possible values for $h$ are chosen from the set $\Phi = \{0.1, 0.25, 0.75, 1.25, 1.75, 2.25, 2.75, 3.25\}$. We continue cycles with the mutations kernel until the median requirement is satisfied empirically by the movement of the particles, or the maximum number of cycles was reached (set at 100). The maximum power of the surrogate first annealing procedure was $\lambda = 0.1$.

To calibrate the surrogate likelihood, we use the transformations described in Section~\ref{sec:cal-approx}. The intercept to account for scaling differences between the log-likelihoods is included in the optimisation, but not used to transform the surrogate log-likelihood as it does not affect the MH ratio. Five-fold cross validation was used to select $\Lambda$ in the lasso procedure.

The SMC sampler was run with the cost of the full likelihood chosen from $L_{F} \in \{0.1, 1, 10, 100, 1000\}$, whilst the cost of the surrogate log-likelihood is fixed at $L_{S} = 0.01$. The relative cost is therefore $\rho \in \{10^1, 10^2, 10^3, 10^4, 10^5, 10^6\}$.

The simulation was repeated 50 times under each setting, from which we measured the average efficiency of the algorithms in two ways. The first efficiency metric was the squared error (SE) of the parameters values multiplied by scaled likelihood evaluations, calculated as $\text{SLE} = E_{L} + \rho^{-1}E_{\tilde{L}}$ where $E_{L}$ is the number of full likelihood evaluations, and $E_{\tilde{L}}$ is the number of surrogate likelihood evaluations. The second efficiency metric used was squared error multiplied by computation time. The median absolute values of the first metric are displayed in Figure~\ref{fig:eff-2000} alongside Tables~\ref{tab:eff-sle}~and~\ref{tab:eff-time} which present the median efficiency gains relative to the standard MH-SMC algorithm under each metric. The raw computation time gains are reported in Table~\ref{tab:comptime}.

\subsection{Simulation results}

The DA+T+SFA algorithm had the best median efficiency gains among all algorithms, tuning methods, and efficiency metrics, with the exception for $\rho = 10^{1}$ on a single occasion, where it was marginally outperformed by DA for the student likelihood under the SE $\times$ SLE metric (Table~\ref{tab:eff-sle}). DA+T+SFA performed best under median tuning for the student-t likelihood and (overall) second to the bootstrap method for the normal likelihood. In the results and tables that follow, we focus on the median method since all tuning methods performed similarly under the normal distribution, and the median method is simpler to program. Also note that the maximum mutation cycle limit (of 100) was never reached for simulations using the median method for the DA+T+SFA simulations.

 Surrogate first annealing with delayed-acceptance and tuning (DA+T+SFA) had efficiency gains ranging from $1.6\times$ to $7.5\times$ (median SE $\times$ SLE, Table~\ref{tab:eff-sle}) and $3.0\times$ to $8.8\times$ (median SE $\times$ time, Table~\ref{tab:eff-time}). The 90th quantile reached $9.1\times$ to $10.5\times$ for $\rho \geq 10^{2}$ when measuring efficiency by SE $\times$ time. DA+T+SFA also had the best median raw computation time improvements, relative to MH-SMC, speeding up the SMC algorithm by $2.9\times$ to $8.8\times$ (see Appendix~\ref{sec:app-sim-fig}, Table~\ref{tab:comptime}).

The best result of the fixed MH-SMC algorithms, MH (f-) or MH (f+), is reported in Tables~\ref{tab:eff-sle}--\ref{tab:comptime} as \textit{MH (fixed)}. These algorithms clearly performed worse than the adaptive MH-SMC with median relative performance (under all metrics) of $0.7\times$ or $0.8\times$. The represents an improvement of at least 25\% for the median tuning method when compared to the ``best'' fixed step-size using MH in SMC (which is not available in practice), adding further evidence that this adaptive procedure is useful for speeding up standard SMC \citep[in line with recommendations from][]{salomone2018unbiased}.

The algorithms using the delayed-acceptance with surrogate likelihood calibration (Section~\ref{sec:cal-approx}, DA + T) outperformed their counterparts without such calibration, in some cases significantly. The exception to this trend occurs for the student-t distribution with the SE $\times$ SLE metric when comparing DA and DA+T. In this scenario, they are mostly on par except for $\rho = 10^{1}$. Less tuning was required for the student-t likelihood (see comparison of posteriors in Appendix~\ref{sec:app-sim-fig}) which may account for this. It is interesting to note that using tuning for the student-t likelihood does increase 10th quantile on average for the efficiency measures when $\rho \geq 10^{3}$.

The efficiency and computation time tables show an interesting feature of the proposed DA+T+SFA algorithm, in that the efficiency gains cannot be solely attributed to either the DA+T aspect or the SFA aspect of the algorithm. Moreover, the efficiency gains for using both DA+T and SFA within the DA+T+SFA algorithm are not additive in its constituent parts.

Given the lighter tails of the normal distribution, we also investigated algorithm performance, on this model, with the maximum annealing parameter of the SFA method set to $\lambda=0.5$ rather than $\lambda = 0.1$. Under this regime, overall speed-ups were observed to be around $0.5\times$ greater than the result reported thus far for the normal model.

From the results, it is clear surrogate likelihoods have the potential to speed-up computation time and efficiency in SMC. However, there does appear to be a threshold for which the ratio of computation cost between the surrogate and full likelihoods must exceed to realise substantial gains. Whilst an $\approx 3\times$ speed-up, as is the case for $\rho = 10^{1}$, may be critical in some cases, it may not justify implementing new methods to improve computation time. Therefore, the DA-SMC methods in this paper are likely to be more valuable when the likelihood cost ratio is closer to $\rho = 10^{2}$ (or greater) and the speed-up can be expected to be about $5.5\times$ to $10.5\times$ (the range of 80\% intervals of the SE $\times$ time metric).

\begin{table}[t]

\caption{\label{tab:eff-sle}Median (80\% interval) multiplicative improvement of efficiency (SE $\times$ SLE) relative to MH-SMC (using median tuning method) for simulation study.}
\centering
\fontsize{10}{12}\selectfont
\begin{tabular}{lll>{\raggedright\arraybackslash}p{6em}>{\raggedright\arraybackslash}p{6em}>{\raggedright\arraybackslash}p{6em}>{\raggedright\arraybackslash}p{6em}>{}p{6em}}
\toprule
Likelihood & Cost ratio ($\rho$) & SFA+DA+T & SFA & DA+T & DA & MH (fixed)\\
\midrule
 & $10^1$ & 1.6 (1.5, 1.7) & 1.5 (1.4, 1.6) & 1.0 (1.0, 1.1) & 1.3 (1.2, 1.4) & 0.7 (0.6, 0.8)\\
\cmidrule{2-7}
 & $10^2$ & 4.2 (3.8, 4.4) & 1.8 (1.7, 2.0) & 2.2 (2.0, 2.4) & 1.5 (1.4, 1.7) & 0.7 (0.7, 0.8)\\
\cmidrule{2-7}
 & $10^3$ & 4.9 (4.6, 5.2) & 1.8 (1.7, 1.9) & 2.5 (2.3, 2.7) & 1.5 (1.4, 1.7) & 0.7 (0.7, 0.8)\\
\cmidrule{2-7}
 & $10^4$ & 5.0 (4.6, 5.5) & 1.8 (1.7, 2.0) & 2.6 (2.3, 2.8) & 1.6 (1.4, 1.7) & 0.7 (0.7, 0.8)\\
\cmidrule{2-7}
 & $10^5$ & 5.1 (4.8, 5.6) & 1.9 (1.7, 2.0) & 2.6 (2.4, 2.8) & 1.6 (1.4, 1.8) & 0.7 (0.7, 0.8)\\
\cmidrule{2-7}
\multirow{-6}{*}{\raggedright\arraybackslash Normal} & $10^6$ & 5.1 (4.6, 5.5) & 1.9 (1.7, 2.0) & 2.6 (2.3, 2.9) & 1.5 (1.5, 1.7) & 0.7 (0.7, 0.8)\\
\cmidrule{1-7}
 & $10^1$ & 2.1 (1.8, 2.3) & 1.7 (1.5, 1.9) & 1.0 (0.9, 1.2) & 2.4 (1.8, 3.1) & 0.8 (0.7, 0.8)\\
\cmidrule{2-7}
 & $10^2$ & 5.6 (4.4, 7.1) & 2.1 (1.8, 2.6) & 2.8 (2.4, 3.2) & 3.2 (2.5, 4.2) & 0.8 (0.7, 0.9)\\
\cmidrule{2-7}
 & $10^3$ & 6.9 (5.6, 8.4) & 2.2 (1.7, 2.6) & 3.1 (2.6, 3.7) & 3.0 (2.1, 3.7) & 0.8 (0.7, 0.8)\\
\cmidrule{2-7}
 & $10^4$ & 7.6 (5.6, 8.6) & 2.3 (1.8, 2.5) & 3.3 (2.7, 4.0) & 3.1 (2.1, 4.6) & 0.8 (0.7, 0.9)\\
\cmidrule{2-7}
 & $10^5$ & 7.1 (5.5, 8.4) & 2.2 (1.8, 2.5) & 3.3 (2.8, 3.8) & 3.3 (2.3, 4.3) & 0.8 (0.7, 0.8)\\
\cmidrule{2-7}
\multirow{-6}{*}{\raggedright\arraybackslash Student} & $10^6$ & 7.5 (6.0, 9.0) & 2.3 (1.8, 2.7) & 3.3 (2.8, 3.9) & 3.2 (2.3, 4.0) & 0.8 (0.7, 0.9)\\
\bottomrule
\end{tabular}
\end{table} 

\begin{table}[t]

\caption{\label{tab:eff-time}Median (80\% interval) multiplicative improvement of efficiency (SE $\times$ Time) relative to MH-SMC (using median tuning method) for simulation study.}
\centering
\fontsize{10}{12}\selectfont
\begin{tabular}{lll>{\raggedright\arraybackslash}p{6em}>{\raggedright\arraybackslash}p{6em}>{\raggedright\arraybackslash}p{6em}>{\raggedright\arraybackslash}p{6em}>{}p{6em}}
\toprule
Likelihood & Cost ratio ($\rho$) & SFA+DA+T & SFA & DA+T & DA & MH (fixed)\\
\midrule
 & $10^1$ & 3.0 (2.7, 3.1) & 1.5 (1.4, 1.6) & 2.4 (2.3, 2.6) & 1.3 (1.2, 1.4) & 0.7 (0.6, 0.8)\\
\cmidrule{2-7}
 & $10^2$ & 5.0 (4.6, 5.4) & 1.8 (1.7, 2.0) & 2.6 (2.4, 2.8) & 1.5 (1.4, 1.7) & 0.7 (0.7, 0.8)\\
\cmidrule{2-7}
 & $10^3$ & 5.3 (5.0, 5.6) & 1.8 (1.7, 1.9) & 2.6 (2.4, 2.8) & 1.5 (1.5, 1.7) & 0.7 (0.7, 0.8)\\
\cmidrule{2-7}
 & $10^4$ & 5.3 (4.9, 5.8) & 1.9 (1.7, 2.0) & 2.7 (2.4, 2.9) & 1.6 (1.4, 1.7) & 0.7 (0.6, 0.8)\\
\cmidrule{2-7}
 & $10^5$ & 5.4 (5.1, 6.0) & 1.9 (1.8, 2.0) & 2.6 (2.5, 2.8) & 1.6 (1.4, 1.8) & 0.7 (0.7, 0.8)\\
\cmidrule{2-7}
\multirow{-6}{*}{\raggedright\arraybackslash Normal} & $10^6$ & 5.4 (4.9, 5.9) & 1.9 (1.8, 2.0) & 2.6 (2.4, 2.9) & 1.6 (1.5, 1.7) & 0.7 (0.7, 0.8)\\
\cmidrule{1-7}
 & $10^1$ & 3.6 (3.0, 3.9) & 1.7 (1.4, 1.9) & 3.2 (2.7, 3.8) & 2.4 (1.8, 3.2) & 0.8 (0.7, 0.8)\\
\cmidrule{2-7}
 & $10^2$ & 7.2 (5.5, 9.1) & 2.2 (1.8, 2.7) & 3.5 (3.1, 4.1) & 3.3 (2.6, 4.4) & 0.8 (0.7, 0.9)\\
\cmidrule{2-7}
 & $10^3$ & 8.1 (6.3, 9.9) & 2.2 (1.7, 2.6) & 3.4 (2.7, 4.0) & 3.2 (2.2, 3.8) & 0.8 (0.7, 0.8)\\
\cmidrule{2-7}
 & $10^4$ & 8.8 (6.2, 10.0) & 2.3 (1.9, 2.6) & 3.5 (2.9, 4.2) & 3.3 (2.2, 4.9) & 0.8 (0.7, 0.9)\\
\cmidrule{2-7}
 & $10^5$ & 8.1 (6.2, 9.7) & 2.2 (1.8, 2.6) & 3.5 (2.9, 4.0) & 3.4 (2.4, 4.6) & 0.8 (0.7, 0.8)\\
\cmidrule{2-7}
\multirow{-6}{*}{\raggedright\arraybackslash Student} & $10^6$ & 8.7 (6.8, 10.5) & 2.4 (1.8, 2.7) & 3.5 (2.9, 4.1) & 3.3 (2.5, 4.2) & 0.8 (0.7, 0.9)\\
\bottomrule
\end{tabular}
\end{table} 

\section{Application with Whittle likelihood}
\label{sec:whittle}

The Whittle likelihood is a computationally efficient likelihood approximation for time series models \citep{whittle1953estimation} constructed using (discrete) Fourier transforms to the frequency domain. A key component of the Whittle likelihood is the periodogram of the series, an estimate of the series' spectral density. The periodogram is asymptotically unbiased, a property inherited by the Whittle likelihood, making it a popular tool in time series modelling. To describe the Whittle likelihood in full, we begin with definitions for a Fourier transform of a time series model's covariance and the discrete Fourier transform of the time series data.

Let $\{X_{t}\}_{t=1}^{n}$ be a zero-mean equally spaced time series with stationary covariance function $\kappa(\tau,\vect{\theta}) = \E(X_{t}X_{t-\tau})$ where $\vect{\theta}$ are parameters of the distribution governing $X_{t}$. Transforming both the data and the covariance function to the frequency domain enables us to construct the Whittle likelihood with these elements rather than using the time domain as inputs. The Fourier transform of the model's covariance function, or the spectral density $f_{\vect{\theta}}(\omega)$, is
\begin{equation*}
  f_{\vect{\theta}}(\omega) = \frac{1}{2\pi} \sum_{\tau = -\infty}^{\infty}\kappa(\tau,\vect{\theta})\exp(-i \omega\tau)
\end{equation*}
where the angular frequency $\omega \in (-\pi, \pi]$. The discrete Fourier transform (DFT) of the time series data is defined as
\begin{equation*}
J(\omega_{k}) = \frac{1}{\sqrt{2\pi}}\sum_{t = 1}^{n}X_{t}\exp(-i \omega_{k}\tau),\quad  \omega_{k} = \frac{2\pi (\left\lceil n/2 \right\rceil + k)}{n}
\end{equation*}
using the Fourier frequencies $\{\omega_{k}\}_{k=1}^{n}$.

The periodogram is an estimate of the spectral density based on the data, and can be calculated using the DFT by
\begin{equation*}
\mathcal{I}(\omega_{k}) = \frac{|J(\omega_{k})|^2}{n}.
\end{equation*}
Using the aforementioned definitions, we can define the Whittle log-likelihood \citep{whittle1953estimation} as
\begin{equation*}
\ell_{\text{whittle}}(\vect\theta) = -\sum_{k = 1}^{n}\left(\log f_{\vect{\theta}}(\omega_{k}) + \frac{\mathcal{I}(\omega_{k})}{f_{\vect{\theta}}(\omega_{k})}\right).
\end{equation*}
In practice the summation over the Fourier frequencies, $\omega_{k}$, need only be evaluated over a subset (less than half) of values due to symmetry about $\omega_{k}=0$ and since $f_{\vect\theta}$(0) = 0.

The periodogram can be calculated in $\mathcal{O}(n \log n)$ time, and only needs to be calculated once per dataset. After dispersing this cost, the cost of each subsequent likelihood evaluation is $\mathcal{O}(n)$, compared to the usual likelihood cost for time series which is $\mathcal{O}(n^2)$.

We demonstrate the use of the Whittle likelihood on an example from \citet{salomone2019spectral} who use the Whittle likelihood for subsampling frequency to obtain a computational efficient MCMC algorithm for long time series data. Our example differs slightly, in that we would like to demonstrate the efficacy of our method in a pre-asymptotic regime. Hence we analyse shorter time series, which can have multimodal posteriors, making them an ideal test for SMC.

We use an autoregressive fractionally integrated moving average model (ARFIMA) to demonstrate the method on a non-trivial model \citep{granger1980introduction}. ARFIMA models are a generalisation of autoregressive integrated moving average model (ARIMA) models using fractional, rather than integer values, of the difference parameter $d$. For~$n$ of adequate length, the zero-mean series $\{X_{t}\}_{t=1}^{n}$ is an ARIFMA time series if
\begin{align*}
\phi(L)\left(1 - L \right)^{d}X_{t} = \theta(L) \varepsilon_{t}
\end{align*}
where $L$ is the lag operator, $\varepsilon_{t}$ is zero-mean Gaussian noise with variance $\sigma^{2}$, and the polynomials $\phi(z)$ and $\theta(z)$ are defined as
\begin{align*}
\phi(z) = 1 - \sum_{i=1}^{p}\phi_{i}z^{i}
\quad\text{ and }\quad \theta(z) = 1 + \sum_{i=1}^{q}\theta_{i}z^{i}.
\end{align*}
These models are fully parameterised by the collection of parameters $(\phi_{1} \cdots \phi_{p})$, $(\theta_{1} \cdots \theta_{q})$, $d$, and $\sigma^{2}$. In order for the ARFIMA process to be stationary, the zeros of $\phi(z)$ and $\theta(z)$ must be outside the complex unit circle, and $-0.5 < d < 0.5$. We impose the stationarity conditions by transforming the polynomial coefficients of $\phi(z)$ and $\theta(z)$ to partial autocorrelations \citep{barndorff1973parametrization}, which only requires that the magnitude of the (transformed) coefficients be less than one for stationarity. Following this transformation, all parameters are mapped to the real line.

ARFIMA models are useful for their ability to describe long-term dependence in time series, which can not be captured by ARIMA models \citep{granger1980introduction}. The spectral density function of an ARIFMA time series is
\begin{equation*}
f_{\vect{\phi},d,\vect{\theta}}(\omega) = \frac{\sigma^2}{2 \pi}\frac{\vert \theta(e^{-i\omega}) \vert^{2}}{\vert \phi(e^{-i\omega}) \vert^{2}} \vert 1 -  e^{-i\omega}\vert^{-2d}
\end{equation*}
as described in \citet[Chapter 7.3.2,][]{brockwell2016introduction}. The spectral density is utilised to calculate the Whittle approximation for the surrogate likelihood.

To test the proposed SMC algorithm we simulated an ARFIMA($p = 2$, $d = 0.4$, $q=1$) time series, of length $8001$, using parameters $\phi_{1} = 0.45$, $\phi_{2} = 0.1$,  $d= 0.4$, $\theta_{1} = -0.4$, and $\sigma^2 = 1$. With $n = 5000$ particles we fit the MH-SMC and DA+T+SFA algorithms, as well as SMC using only the surrogate likelihood (labelled \textit{surrogate only}). The surrogate likelihood was calibrated using the transformation described in Section~\ref{sec:cal-approx} without a shift transformation as the surrogate and full likelihoods are well aligned. The maximum power of the surrogate first annealing procedure was $\lambda = 0.01$ which reflects the high number of observations. The median tuning method was used to select the optimal step size, and the mutation step of the SMC algorithm ran until the median empirical total ESJD was greater than 3.

When calculating the estimate of the covariance matrix, $\Sigma$, for using in these algorithms it was necessary to demean the particles locations with respect to the mode they occupied. This avoided a close to singular estimate for $\Sigma$, and better reflected the average local covariance structure about the modes. We estimated the modes with k-means clustering \citep{hartigan1979algorithm} where the number of clusters was selected using the Duda-Hart test (\citeyear{duda1973pattern}) then Calinski-Harabasz criterion (\citeyear{calinski1974dendrite}) as implemented in the package \texttt{fpc} \citep{hennigfpc} in \texttt{R}.

Overall, the results held a $3.9\times$ to $5.8\times$ speed-up across the 10 simulations (80\% interval). In real terms, this reduced the computation time from about 20.5 hours to 4.5 hours. The full likelihood was evaluated $5.3\times$ to $7.7\times$ more often during the standard MH-SMC algorithm, as compared to the DA+T+SFA version. The cost ratio of the full likelihood to Whittle likelihood was approximately $10^{2.75}$ to $10^3$, indicating that the speeds up were in line with the simulation study in Section~\ref{sec:sim-ex}, but slightly less than expected.

\begin{figure}
    \centering
    \includegraphics[width = 0.9\textwidth]{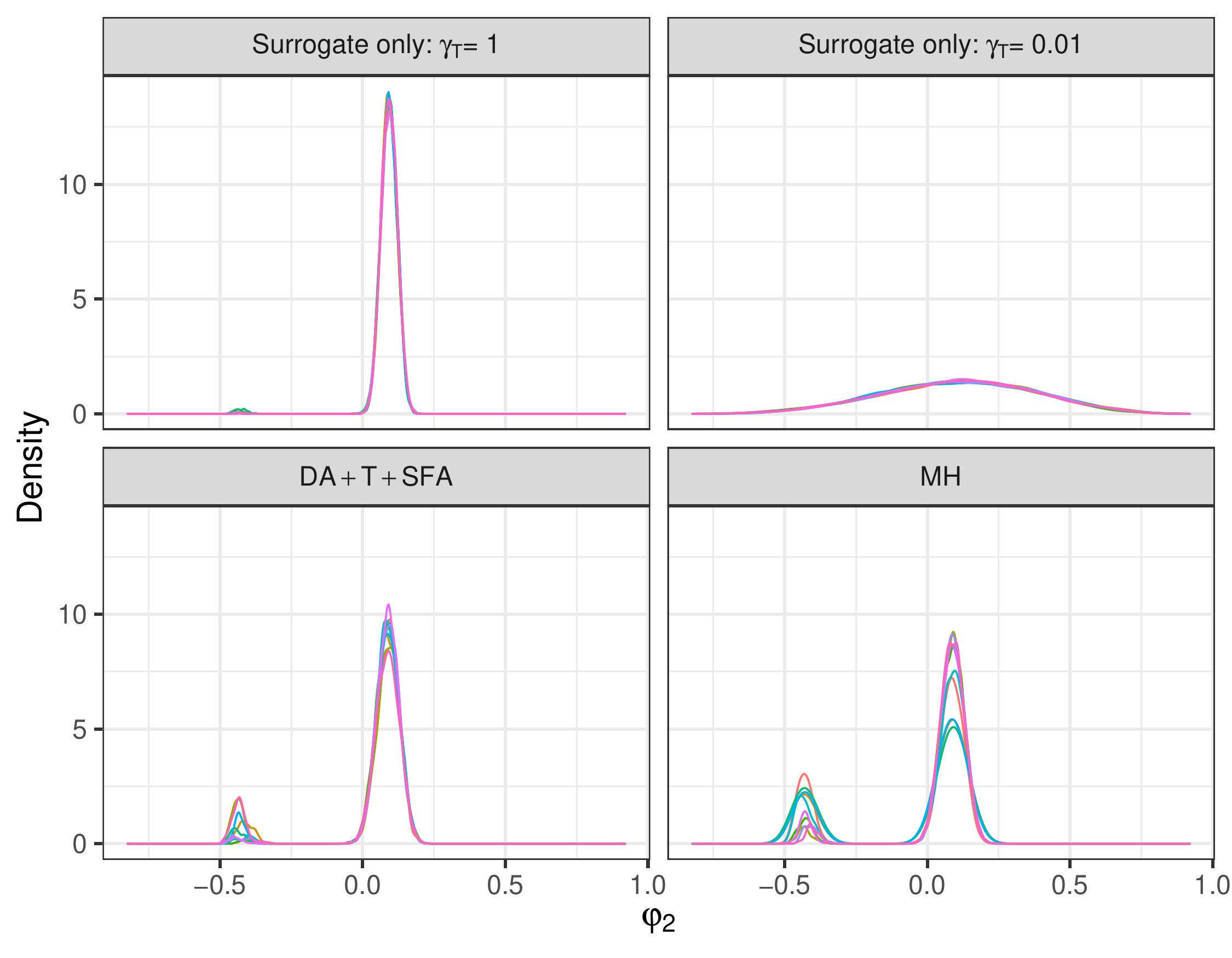}
    \caption{Posteriors of $\phi_{2}$ from 10 replicates of four SMC algorithms. The SMC algorithms using only the surrogate likelihood (approx) are annealed to $\gamma_{T} \in \{0.01,1\}$. The latter of which is the initial particle set for the surrogate first annealing procedure.}
    \label{fig:whittle-phi2}
\end{figure}

Figure~\ref{fig:whittle-phi2} displays the density plots of $\phi_{2}$ from the SMC algorithms over the replicates, whilst the remaining densities are displayed in Appendix~\ref{sec:app-whittle-fig}. These density comparisons demonstrate that using surrogate likelihoods in SMC can be done with minimal accuracy lost in the posterior computations, with an appreciable gain in speed. This was a challenging model to consider because the multimodality in the full posterior is not well approximated by the surrogate posterior. Comparing the top-left facet of Figure~\ref{fig:whittle-phi2} to the bottom-right facet illustrates this.

In an additional experiment, where the length of the time series was 10001 (rather than 8001), we observed speed-ups of $4.9\times$ to $6.4\times$. Under this particular simulated dataset, the target posterior did not exhibit multimodality which may partially explain the increase. The adaptive SMC methods proposed in this paper are able to perform well under both cases, with and without multimodality.

\section{Discussion}
\label{sec:con}

We have explored several ways of using surrogate likelihoods to improve the efficiency of SMC. In particular, delayed-acceptance within the mutation step with calibration using the population of particles, and surrogate first annealing, were proposed and used to this end.

A prevailing assumption of ours has been that a surrogate likelihood is available for the application at hand. In the absence of a good candidate it may be convenient to use a variational Bayes approximation as a surrogate posterior \citep[see for example][Ch. 10]{bishop2006pattern}. In this case the surrogate first annealing method would be similar to \citet{donnet2017using}, who start with a variational approximation as their initial distribution.  Non-parametric surrogate likelihoods could also be considered, such as nearest-neighbour or Gaussian Processes \citep[see for example,][]{sherlock2017adaptive,drovandi2018accelerating}.

A computational aspect of delayed-acceptance, particularly important in SMC, is its effect on parallel computation. The mutation step of SMC can be easily parallelised but delayed-acceptance results in some mutations occurring quickly (first stage rejection), whilst others taking considerably longer. Whilst we did not observe any adverse behaviour our settings, appropriate scheduling for parallel implementations should be considered.

The SFA method is sensitive to the choice of surrogate temperature, $\lambda$, especially if there is a large mismatch between the surrogate and full likelihoods. In this case, if $\lambda$ is too low, then computational efficiency is lost, but if $\lambda$ is too high, the final posterior can be inaccurate. In the simulation study and ARFIMA example, $\lambda$ was chosen by inspecting the surrogate likelihood and determining a value for $\lambda$ which retained sufficient density to approximately cover the two peaks observed from the full likelihood. In general, more observations would preclude a lower value of $\lambda$. Future work in determining an appropriate $\lambda$ automatically would be a useful contribution.

The theory contributed by this paper proposes a general framework for choosing tuning parameters in SMC, with a focus on the typically costly mutation step. The tuning parameter decision is cast as an optimisation problem of cost minimisation, subject to a sufficient quality of diversification. The framework connects with and generalises, several tuning methods in the SMC literature, allowing them to be used with delayed-acceptance kernels. We have provided evidence that this framework is appropriate for improving computational efficiency with both Metropolis-Hastings and delayed-acceptance kernels in SMC without burdensome input from the user.

\bibliographystyle{agsm}
\bibliography{references}  

\appendix

\section{Appendix}

\subsection{One-move diversification}\label{sec:mad}

An alternative to ESJD diversification can be found in a simple method from \citet{south2019seq} for choosing the number of MCMC runs --- the basis of which is from \citet{drovandi2011likelihood}. In this regime the number of cycles, $k$, is chosen so that each particle moves at least once in $k$ iterations. A move occurs when a proposal is accepted using an MH kernel.  We will refer to this criterion as one-move diversification. For a fixed scaling parameter $h$, one-move diversification uses the MH acceptance rates to determine the average number of MCMC cycles required for at least one proposal per particle to be accepted. More generally, one could require a higher minimum number of moves, but for simplicity we just consider the case of at least one move.

This section will consider a single mutation step of the SMC algorithm, consisting of multiple cycles of the MCMC kernel, indexed by $s \in \{1,2, \ldots, k\}$. Assuming the probability of moving (or acceptance, $\alpha^{(1)}$) is equal across cycles, the average probability (across the tempered posterior distribution) that at least one is accepted in a sequence of $k$ cycles,~$\alpha^{(k)}$, is
\begin{align}
\alpha^{(k)}  = 1 - \left(1-\alpha^{(1)}\right)^{k} \label{eq:alpha-k}
\end{align}
for $k \in \{1,2,\ldots\}$. A pilot mutation step can be used to estimate the average acceptance rate across the particles in a single step, $\widehat{\alpha}^{(1)}$. We can then find $k$ such that $\alpha^{(k)} \geq p_{\min}$ for some threshold $0 < p_{\min} < 1$. The formula to choose the total number of iterations, $k$, is
\begin{align}
k = \left\lceil \frac{\log(1 - p_{\min})}{\log\left(1-\widehat{\alpha}^{(1)}\right)} \right\rceil \label{eq:MADsimple}
\end{align}
where $\widehat{\alpha}^{(1)}$ is the estimated acceptance rate from the pilot run of the MH kernel.

To frame this in the context of optimising computation time, note that the underlying criterion is to ensure a sufficient number of mutation steps are taken so that the probability of at least one move is greater than $p_{\min}$ for a given particle.

If we denote a move by $\Vert \vect{\theta}_{s} -  \vect{\theta}_{s-1}\Vert_{0}$, where $\Vert \cdot \Vert_{0}$ is the zero ``norm'', the corresponding diversification criterion can be expressed with
\begin{align}
  D(k, \vect{\phi}) = \Pr\left(\sum_{s=1}^{k} \left\Vert \vect{\theta}_{s} -  \vect{\theta}_{s-1} \right\Vert_{0} \geq 1 \right) \quad \text{and} \quad d=  p_{\min}
    \label{eq:div-mad-gen}
\end{align}
where the probability is taken with respect to the acceptance rates of the Metropolis-Hastings steps. Of course, this expression for $D(k, \vect{\phi})$ is a more general version of \eqref{eq:alpha-k} and coincides if we assume the probability of acceptance is equal across particle locations and MCMC iterations, $s$. We emphasise the norm notation to draw a comparison to the jumping distance diversification in Section~\ref{sec:jdd-std}. That is, we can write $P(k, \vect{\phi})$ in~\eqref{eq:div-jdd-gen} as
\begin{equation*}
P(k, \vect{\phi}) = \Pr\left( \E \left[\sum_{s=1}^{k} \left\Vert \vect{\theta}_{s} -  \vect{\theta}_{s-1} \right\Vert^{2}_{\Sigma} \right] \geq d \right)
\end{equation*}
where the expectation is with respect to the random acceptance over $k$ cycles of the MH kernel. Written in this way, $P(k, \vect{\phi})$ elicits an interesting comparison to \eqref{eq:div-mad-gen}; it is a change of ``norm'' when moving between one-move and jumping distance diversification.

Now we wish to use one-move criterion to select the tuning parameters. If we use different proposal kernel tuning parameters for particular subsets of particles, the acceptance rate will be a function of those parameters, so we write $\alpha^{(k)}$ as $\alpha^{(k)}(\vect{\phi})$. The optimisation stated in \eqref{eq:generalopt} can be simplified as stated in Proposition~\ref{th:best-om}.
\begin{prop}\label{th:best-om}
	Assume the cost function is $C(k, \vect{\phi}) = k \times L_{F}$, approximating the cost of a standard Metropolis-Hastings step, and $D(k,\vect{\phi}) = \alpha^{(k)}(\vect{\phi})$. The latter also corresponds to \eqref{eq:div-mad-gen} assuming a uniform acceptance rate across the support of $\vect{\theta}$.  Then the general problem in \eqref{eq:generalopt} is equivalent to
\begin{align}
    \argmin_{\vect{\phi} \in \Phi}~   \frac{\log(1 - p_{\min})}{\log\left(1-\alpha^{(1)}(\vect{\phi})\right)}  \label{eq:MADopt}
\end{align}
where the general diversification threshold, $d$, has been replaced by the probability $p_{\min}$.
\end{prop}

Proposition~\ref{th:best-om} is the solution to choosing the best tuning parameters with the one-move criterion and MH-cost. It closely connects to the original decision for $k$ without tuning parameters \eqref{eq:MADsimple}. A proof of Proposition~\ref{th:best-om} is in Appendix~\ref{pr:best-om}.

In general, we expect the tuning criterion in Proposition~\ref{th:best-om} to perform poorly. This can be demonstrated by a simple, but highly applicable, example. If the tuning parameter is the step size for an MH mutation, i.e. $\vect{\phi} = [h]$, then we would expect the acceptance probability, $\alpha^{(1)}(\vect{\phi})$, to be monotone decreasing in $h$. Hence the minimisation in \eqref{eq:MADopt} will prefer the minimum step size possible, which will ensure at least one move with the minimal computation cost. In other words, the diversification criterion in \eqref{eq:div-mad-gen} is only concerned with the probability of at least one move, not the quality of this move.

Due to the aforementioned shortcoming, a diversification criterion that also measures the quality of the mutation is desirable. For this reason, we focus on the ESJD as a criterion in the main text.

\subsection{Proof of Proposition~\ref{th:esjd-jen}}\label{pr:esjd-jen}
Let $\mathsf{D}_m = \left\{(k,\vect{\phi}) \in \mathbb{Z}^{+} \times \Phi:D(k,\vect{\phi}) \geq d\right\}$ with $D(k,\vect{\phi}) = \med\left\{\sum_{s=1}^{k} J_{s}(\vect{\phi})  \right\}$. Using the multivariate Jensen inequality for medians in \citet[][Theorem 5.2]{merkle2010jensen} we have that
\begin{equation*}
  \sum_{s=1}^{k}\med\left\{ J_{s}(\vect{\phi})  \right\} \leq \med\left\{\sum_{s=1}^{k} J_{s}(\vect{\phi})  \right\}
\end{equation*}
and assuming the jumping distances are iid, for a given $\vect{\phi}$, we further reduce this to
\begin{equation}
  k  \times \med\left\{ J_{1}(\vect{\phi})  \right\} \leq \med\left\{\sum_{s=1}^{k} J_{s}(\vect{\phi})  \right\} \label{eq:multi-med-ineq}.
\end{equation}
We can define the set $\tilde{\mathsf{D}}_m$ as
\begin{gather*}
    \tilde{\mathsf{D}}_m = \left\{(k,\vect{\phi}) \in \mathbb{Z}^{+} \times \Phi:\tilde{D}(k,\vect{\phi}) \geq d\right\} \\
    \tilde{D}(k, \vect{\phi}) =k  \times \med\left\{ J_{1}(\vect{\phi}) \right\}
\end{gather*}
then we see from \eqref{eq:multi-med-ineq} that $\tilde{\mathsf{D}}_m \subseteq \mathsf{D}_m$.

\subsection{Proof of Proposition~\ref{th:esjd-mh}}\label{pr:esjd-mh}

Under the MH cost function, $C(k, \vect{\phi}) = k \times L_{F}$, and approximate ESJD diversification criterion,
\begin{align*}
\tilde{\mathsf{D}}_m = &\left\{ (k,\vect{\phi}) \in \mathbb{Z}^{+} \times \Phi: \tilde{D}(k,\vect{\phi}) \geq d\right\}\\
\text{where } &\tilde{D}(k,\vect{\phi}) = k \times \med\left\{ J_{1}(\vect{\phi}) \right\},
\end{align*}
the inequality for the diversification criterion can be rearranged into
\begin{equation*}
	k \geq \frac{d}{\med\left\{ J_{1}(\vect{\phi}) \right\}}.
\end{equation*}
Under this restriction, note that
\begin{equation*}
	C(k, \vect{\phi}) \geq L_{F} \times \frac{d}{\med\left\{ J_{1}(\vect{\phi}) \right\}}.
\end{equation*}
so under these conditions, the general problem in \eqref{eq:generalopt} is equivalent to
\begin{align*}
    \argmin_{\vect{\phi} \in \Phi}~  \left( \med\left\{ J_{1}(\vect{\phi}) \right\} \right)^{-1} \equiv \argmax_{\vect{\phi} \in \Phi}~  \med\left\{ J_{1}(\vect{\phi})  \right\}.
\end{align*}

\subsection{Proof of Proposition~\ref{th:best-om}}\label{pr:best-om}

Under the MH cost function, $C(k, \vect{\phi}) = k \times L_{F}$, and one-move diversification criterion,
\begin{align*}
\mathsf{D} = &\left\{ (k,\vect{\phi}) \in \mathbb{Z}^{+} \times \Phi: \alpha^{(k)}(\vect{\phi}) \geq p_{\min}\right\}\\
\text{where } &\alpha^{(k)}(\vect{\phi}) = 1 - (1-\alpha^{(1)}(\vect{\phi}))^{k},
\end{align*}
the inequality for the diversification criterion can be rearranged into
\begin{equation*}
	k \geq \frac{\log(1 - p_{\min})}{\log(1-\alpha^{(1)}(\vect{\phi}))}.
\end{equation*}
Under this restriction, note that
\begin{equation*}
	C(k, \vect{\phi}) \geq L_{F} \times \frac{\log(1 - p_{\min})}{\log(1-\alpha^{(1)}(\vect{\phi}))}.
\end{equation*}
so  under these conditions, the general problem in \eqref{eq:generalopt} is equivalent to
\begin{align*}
    \argmin_{\vect{\phi} \in \Phi}~   \frac{\log(1 - p_{\min})}{\log(1-\alpha^{(1)}(\vect{\phi}))}.
\end{align*}

\clearpage
\newpage

\subsection{Additional figures and tables from simulation}\label{sec:app-sim-fig}

\begin{table}[htp]

\caption{\label{tab:comptime}Median (80\% interval) multiplicative improvement of computation time relative to MH-SMC (using median tuning method) for simulation study.}
\centering
\fontsize{10}{12}\selectfont
\begin{tabular}{lll>{\raggedright\arraybackslash}p{6em}>{\raggedright\arraybackslash}p{6em}>{\raggedright\arraybackslash}p{6em}>{\raggedright\arraybackslash}p{6em}>{}p{6em}}
\toprule
Likelihood & Cost ratio ($\rho$) & SFA+DA+T & SFA & DA+T & DA & MH (fixed)\\
\midrule
 & $10^1$ & 2.9 (2.8, 3.0) & 1.5 (1.5, 1.6) & 2.4 (2.4, 2.6) & 1.3 (1.3, 1.4) & 0.7 (0.7, 0.8)\\
\cmidrule{2-7}
 & $10^2$ & 5.0 (4.8, 5.2) & 1.8 (1.8, 1.8) & 2.6 (2.5, 2.7) & 1.5 (1.5, 1.6) & 0.7 (0.7, 0.7)\\
\cmidrule{2-7}
 & $10^3$ & 5.3 (5.1, 5.6) & 1.9 (1.9, 2.0) & 2.6 (2.5, 2.8) & 1.6 (1.5, 1.6) & 0.7 (0.7, 0.8)\\
\cmidrule{2-7}
 & $10^4$ & 5.4 (5.1, 5.7) & 1.9 (1.9, 1.9) & 2.6 (2.5, 2.8) & 1.6 (1.5, 1.6) & 0.7 (0.7, 0.7)\\
\cmidrule{2-7}
 & $10^5$ & 5.4 (5.1, 5.6) & 1.9 (1.9, 1.9) & 2.6 (2.5, 2.7) & 1.6 (1.5, 1.6) & 0.7 (0.7, 0.7)\\
\cmidrule{2-7}
\multirow{-6}{*}{\raggedright\arraybackslash Normal} & $10^6$ & 5.5 (5.2, 6.0) & 1.9 (1.9, 2.0) & 2.6 (2.5, 2.8) & 1.6 (1.5, 1.7) & 0.7 (0.7, 0.8)\\
\cmidrule{1-7}
 & $10^1$ & 3.5 (3.2, 3.9) & 1.7 (1.5, 1.9) & 3.2 (3.0, 3.7) & 2.5 (2.1, 2.8) & 0.8 (0.7, 0.8)\\
\cmidrule{2-7}
 & $10^2$ & 7.5 (5.7, 8.5) & 2.2 (1.8, 2.5) & 3.5 (3.1, 4.0) & 3.3 (2.7, 3.9) & 0.8 (0.7, 0.8)\\
\cmidrule{2-7}
 & $10^3$ & 8.2 (6.4, 9.7) & 2.2 (1.8, 2.6) & 3.5 (3.0, 3.9) & 3.4 (2.7, 4.0) & 0.8 (0.7, 0.8)\\
\cmidrule{2-7}
 & $10^4$ & 8.6 (6.6, 9.7) & 2.3 (1.8, 2.6) & 3.6 (3.2, 4.0) & 3.5 (2.6, 4.4) & 0.8 (0.7, 0.8)\\
\cmidrule{2-7}
 & $10^5$ & 8.1 (6.4, 9.4) & 2.2 (1.8, 2.6) & 3.5 (3.1, 3.8) & 3.4 (2.7, 4.2) & 0.7 (0.7, 0.8)\\
\cmidrule{2-7}
\multirow{-6}{*}{\raggedright\arraybackslash Student} & $10^6$ & 8.8 (7.3, 9.8) & 2.3 (1.9, 2.6) & 3.6 (3.2, 4.0) & 3.6 (2.8, 4.0) & 0.8 (0.7, 0.8)\\
\bottomrule
\end{tabular}
\end{table}

\begin{figure}[htp]
    \centering
    \includegraphics[width = 0.9\textwidth]{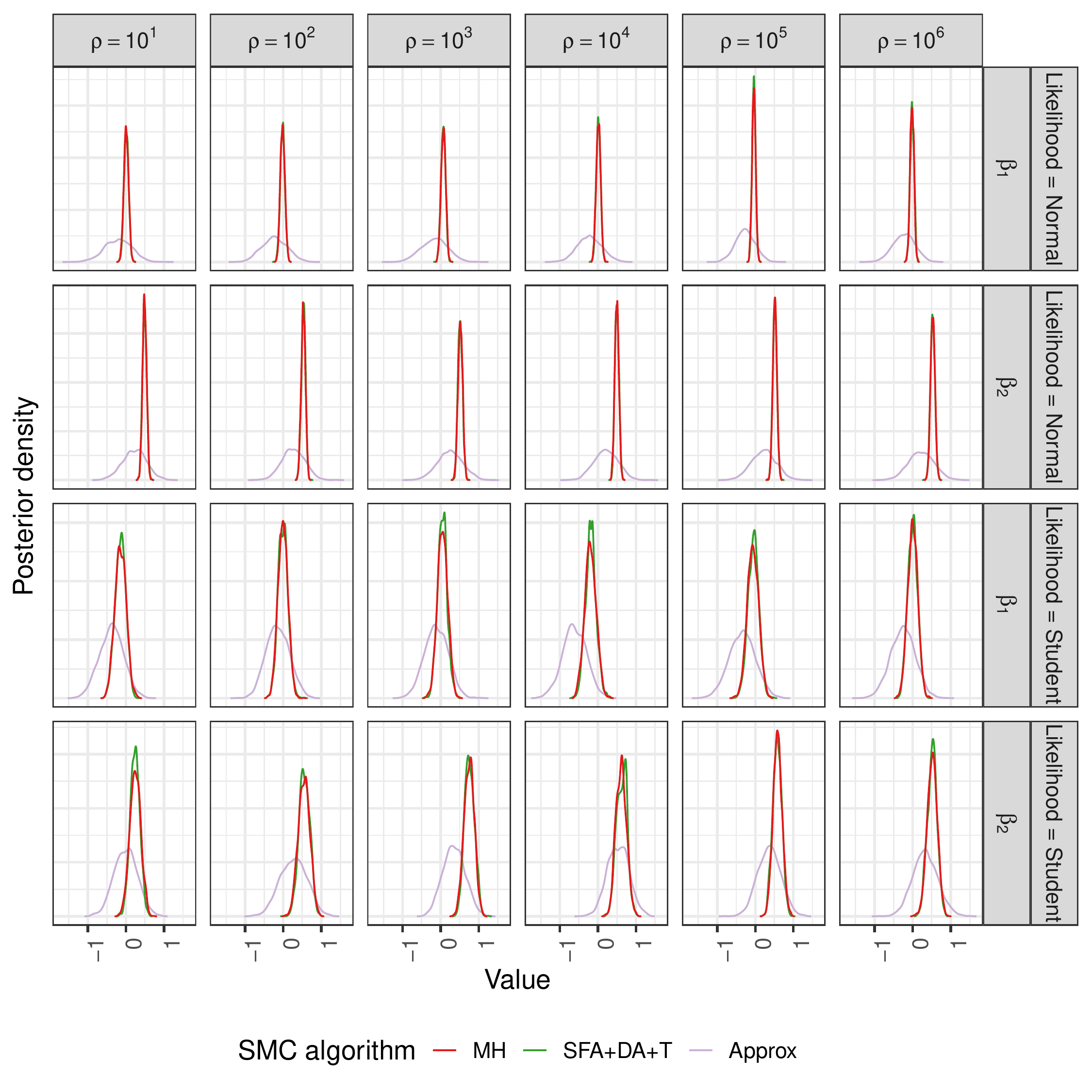}
    \caption{Example of posterior $\vect{\beta}$ densities from three SMC algorithms (example 1).}
    \label{fig:beta-dens-1}
\end{figure}

\begin{figure}[htp]
    \centering
    \includegraphics[width = 0.9\textwidth]{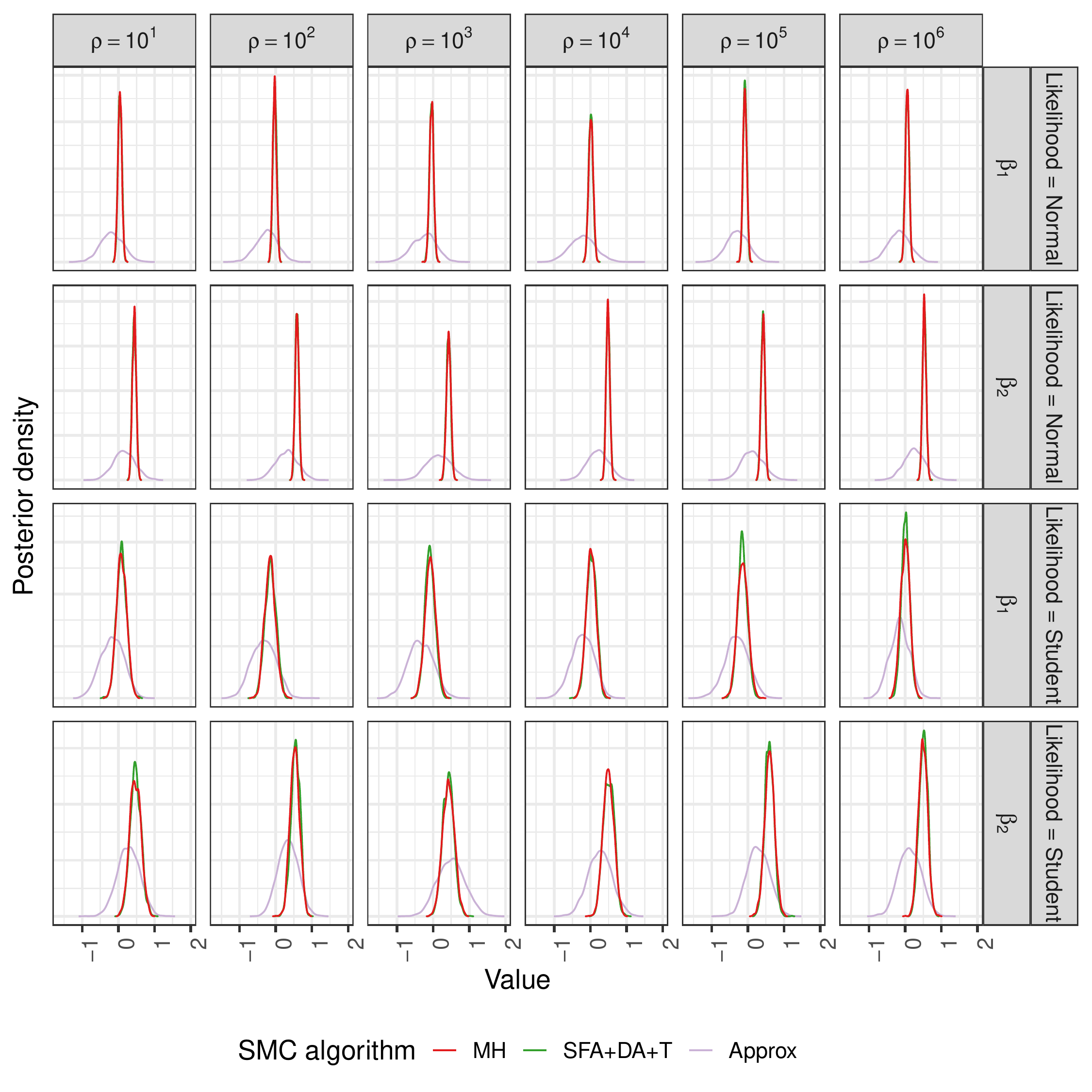}
    \caption{Example of posterior $\vect{\beta}$ densities from three SMC algorithms (example 2).}
    \label{fig:beta-dens-2}
\end{figure}

\begin{figure}[htp]
    \centering
    \includegraphics[width = 0.9\textwidth]{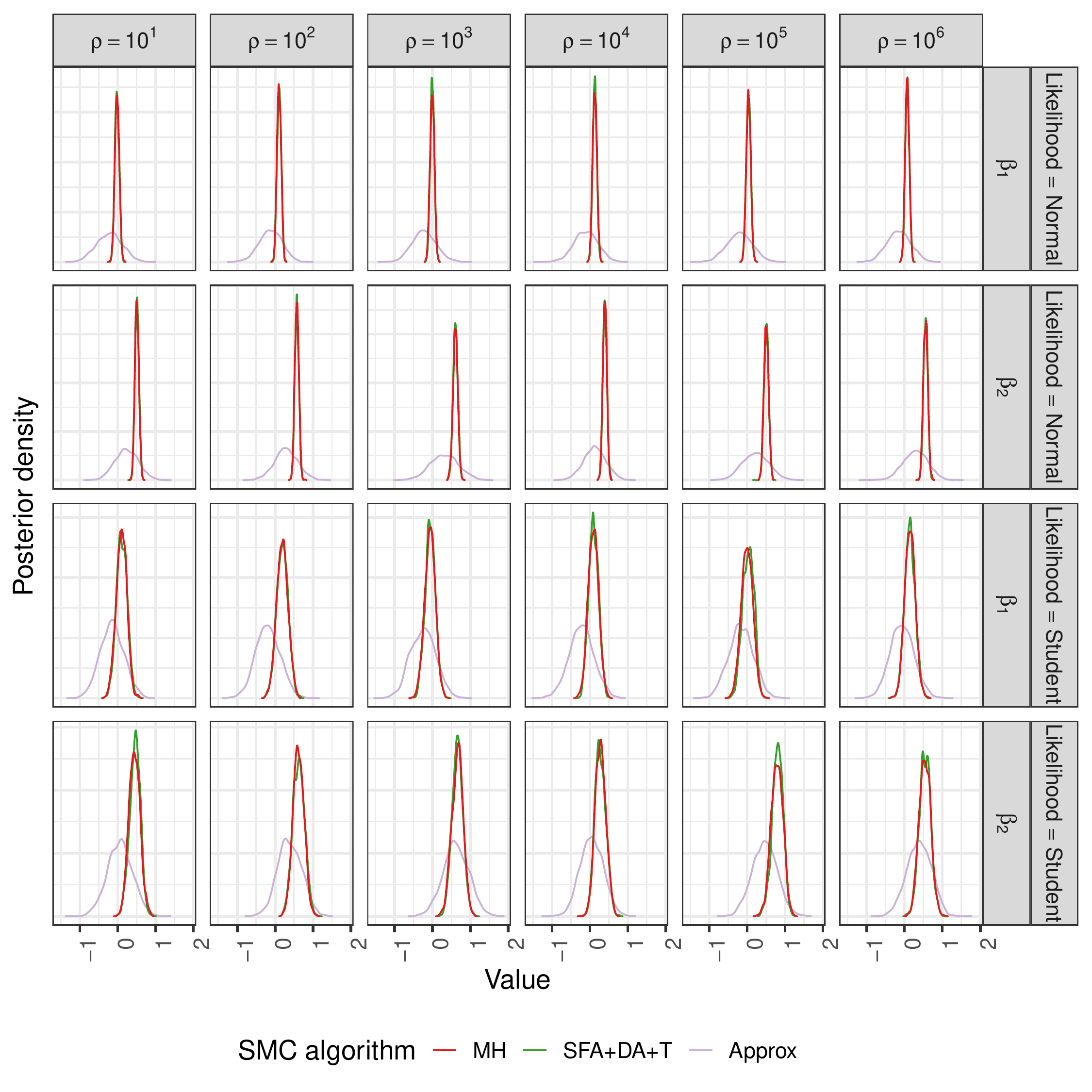}
    \caption{Example of posterior $\vect{\beta}$ densities from three SMC algorithms (example 3).}
    \label{fig:beta-dens-3}
\end{figure}

\begin{figure}[htp]
    \centering
    \includegraphics[width = 0.9\textwidth]{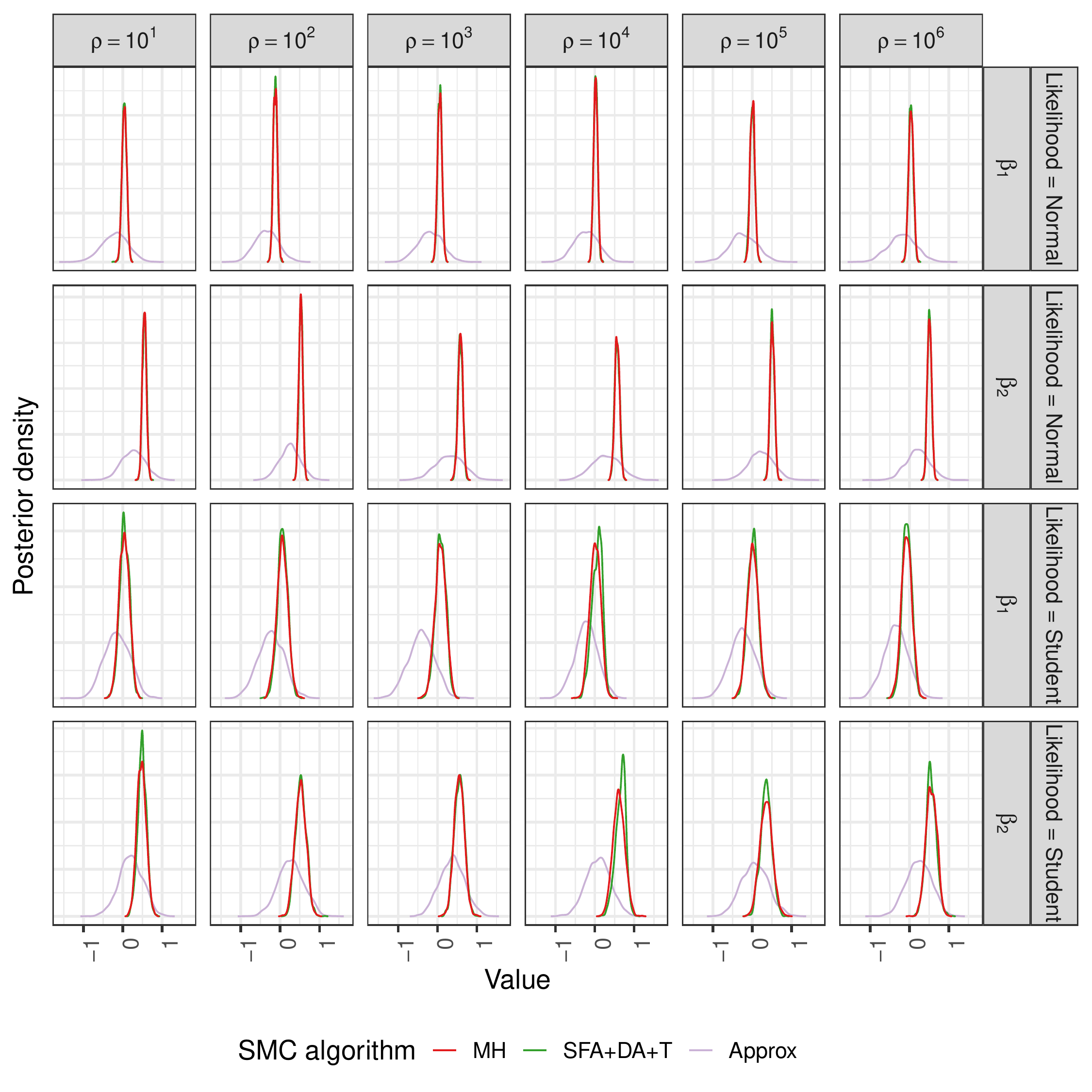}
    \caption{Example of posterior $\vect{\beta}$ densities from three SMC algorithms (example 4).}
    \label{fig:beta-dens-4}
\end{figure}

\clearpage
\newpage

\subsection{Additional figures from ARFIMA model example}\label{sec:app-whittle-fig}

\begin{figure}[htp]
    \centering
    \includegraphics[width = 0.9\textwidth]{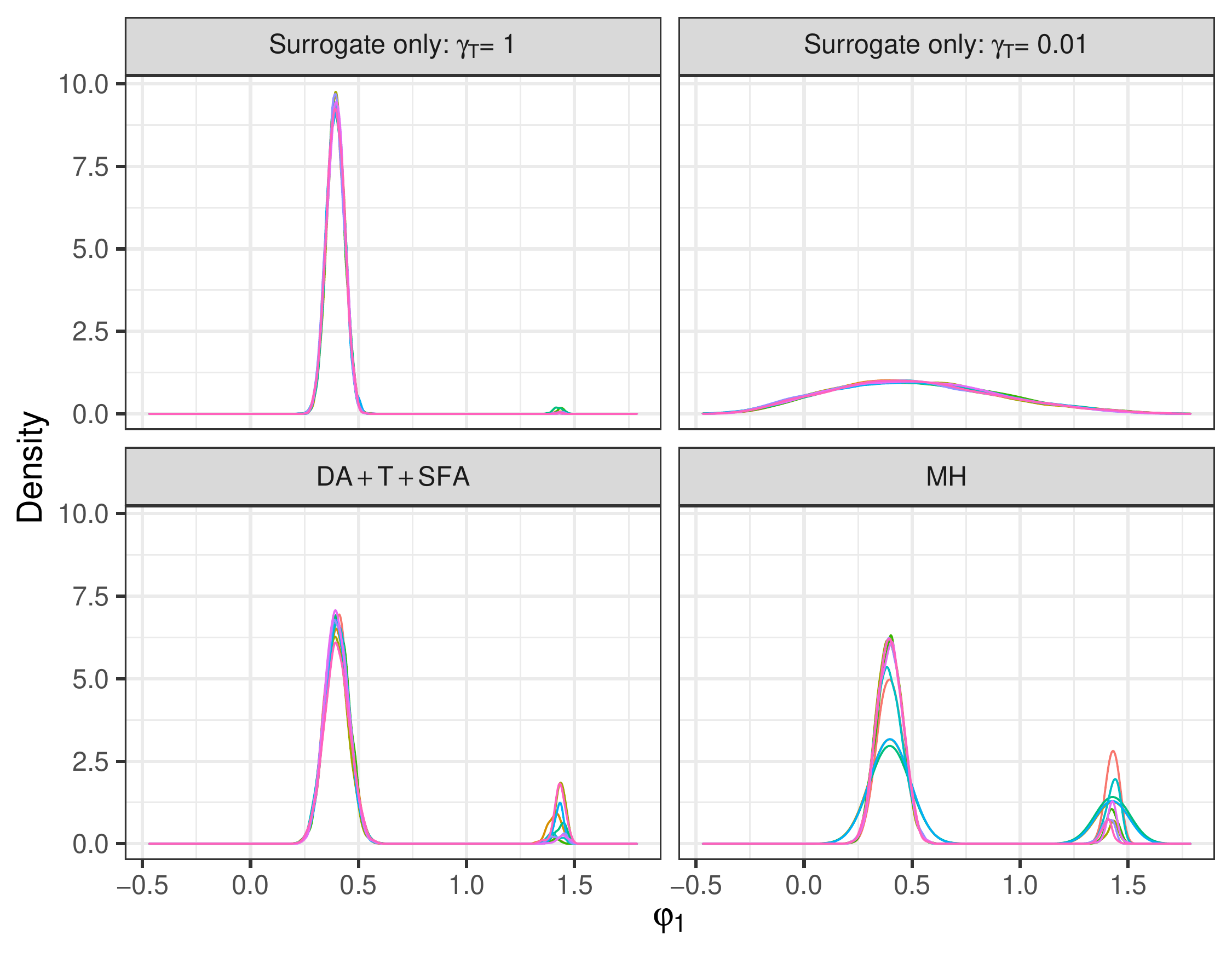}
    \caption{Posteriors of $\phi_{1}$ from 10 replicates of the four SMC algorithms. The SMC algorithms using only the surrogate likelihood (approx) are annealed to $\gamma_{T} \in \{0.01,1\}$ and $\gamma_{T} = 0.01$. The latter of which is the initial particle set for the surrogate first annealing procedure.}
    \label{fig:whittle-phi1}
\end{figure}

\begin{figure}[htp]
    \centering
    \includegraphics[width = 0.9\textwidth]{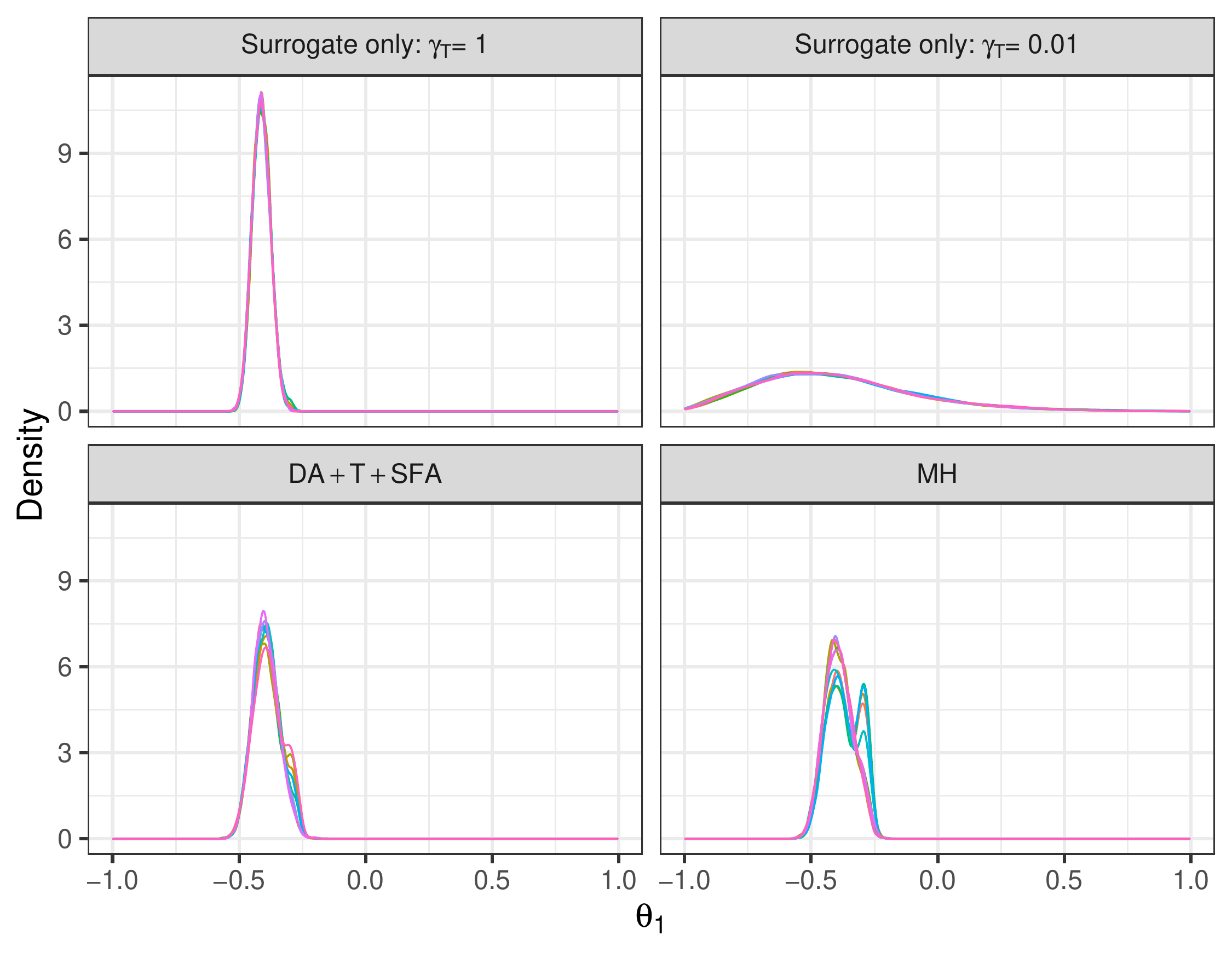}
    \caption{Posteriors of $\theta_{1}$ from 10 replicates of the four SMC algorithms. The SMC algorithms using only the surrogate likelihood (approx) are annealed to $\gamma_{T} \in \{0.01,1\}$ and $\gamma_{T} = 0.01$. The latter of which is the initial particle set for the surrogate first annealing procedure.}
    \label{fig:whittle-theta}
\end{figure}

\begin{figure}[htp]
    \centering
    \includegraphics[width = 0.9\textwidth]{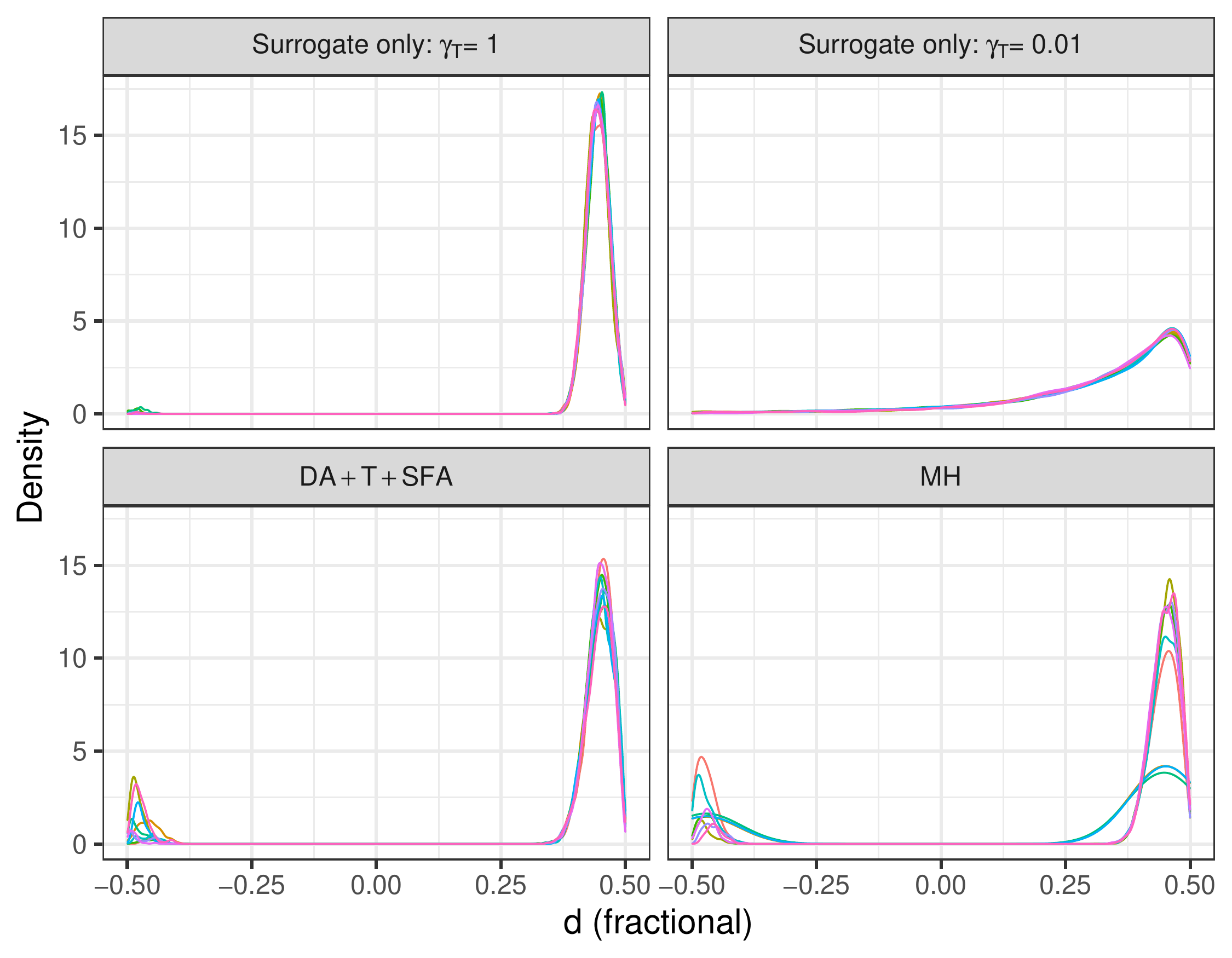}
    \caption{Posteriors of $d$ from 10 replicates of the four SMC algorithms. The SMC algorithms using only the surrogate likelihood (approx) are annealed to $\gamma_{T} \in \{0.01,1\}$ and $\gamma_{T} = 0.01$. The latter of which is the initial particle set for the surrogate first annealing procedure.}
    \label{fig:whittle-dfrac}
\end{figure}

\begin{figure}[htp]
    \centering
    \includegraphics[width = 0.9\textwidth]{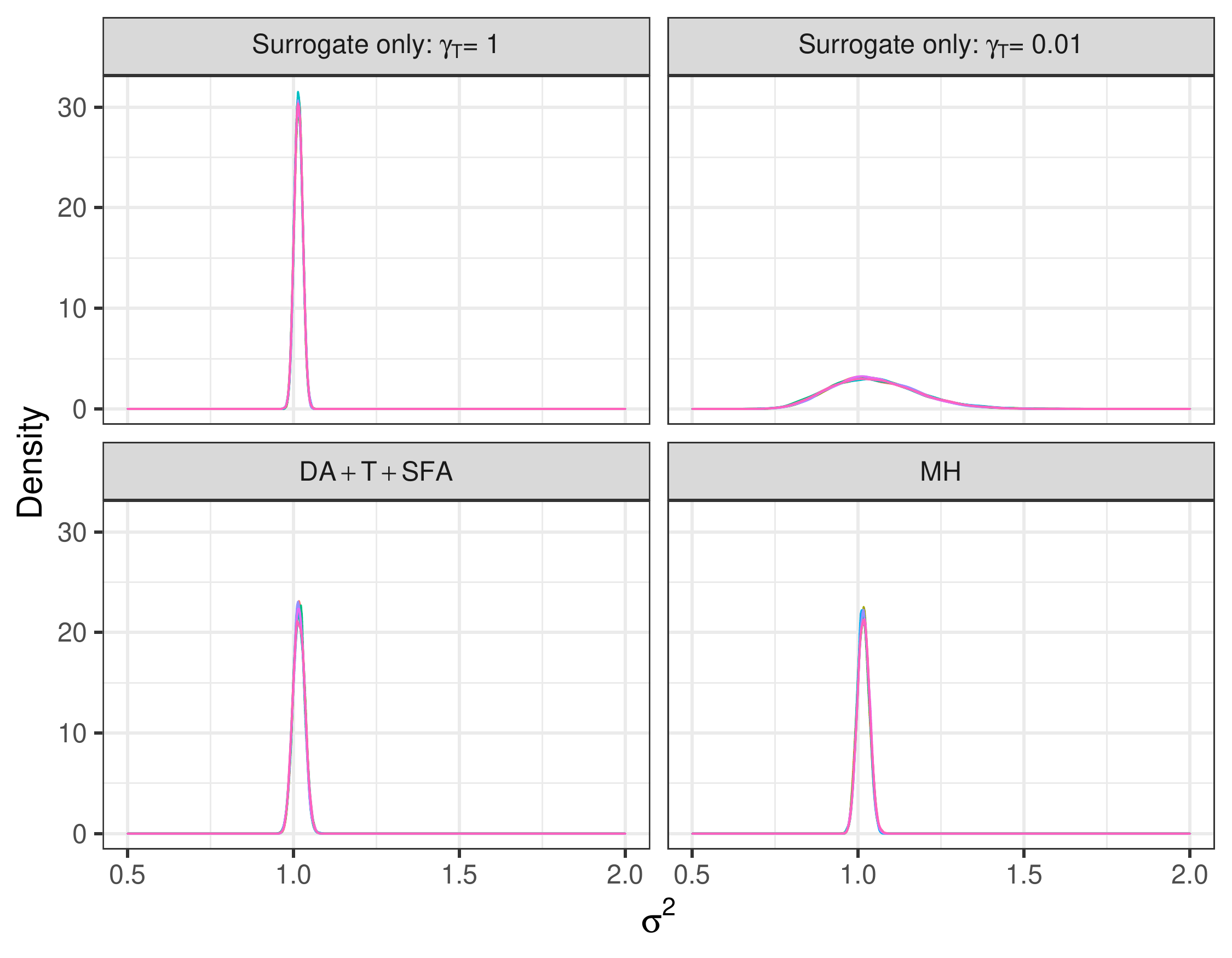}
    \caption{Posteriors of $\sigma^{2}$ from 10 replicates of the four SMC algorithms. The SMC algorithms using only the surrogate likelihood (approx) are annealed to $\gamma_{T} \in \{0.01,1\}$ and $\gamma_{T} = 0.01$. The latter of which is the initial particle set for the surrogate first annealing procedure.}
    \label{fig:whittle-sigma2}
\end{figure}

\end{document}